\documentclass[11pt,a4paper]{article}
\usepackage{jheppub}
\usepackage{natbib}
\usepackage{graphicx}
\usepackage[dvipsnames]{xcolor}

\usepackage{mathtools}
\usepackage{hyperref}
\usepackage{float}
\usepackage{comment}
\usepackage{multirow}
\usepackage{indentfirst}

\def\bea{\begin{eqnarray} }
\def\eea{ \end{eqnarray} } 

\definecolor{Blu}{rgb}{0.,0.,1.}

\newcommand{\Mp}{M_{PL}}
\newcommand{\rhoBBNT}{\rho_{\phi}(T_{\rm{BBN}})}

\author[a,b]{Anirban Biswas,}
\author[b]{Arpan Kar,}
\author[b,c]{Bum-Hoon Lee,}
\author[b,c]{Hocheol Lee,}
\author[b]{Wonwoo Lee,}
\author[b,c]{Stefano Scopel,}
\author[b,c]{Liliana Velasco-Sevilla,}
\author[b,d]{Lu Yin}
\affiliation[a]{Department of Physics \& Lab of Dark Universe, Yonsei University, Seoul 03722, Republic of Korea}
\affiliation[b]{Center for Quantum Spacetime, Sogang University, Seoul 121-742, South Korea}
\affiliation[c]{Department of Physics, Sogang University, Seoul 121-742, South Korea}
\affiliation[d]{Asia Pacific Center for Theoretical Physics (APCTP)
San 31, Hyoja-dong, Nam-gu, Pohang 790-784, South Korea}
\emailAdd{anirban.biswas.sinp@gmail.com}
\emailAdd{arpankarphys@gmail.com}
\emailAdd{bhl@sogang.ac.kr}
\emailAdd{insaying@sogang.ac.kr}
\emailAdd{warrior@sogang.ac.kr}
\emailAdd{scopel@sogang.ac.kr}
\emailAdd{liliana.velascosevilla@gmail.com}
\emailAdd{lu.yin@apctp.org}

\title{WIMPs in Dilatonic Einstein Gauss-Bonnet Cosmology}

\abstract{We use the Weakly Interacting Massive Particle (WIMP) thermal decoupling scenario to probe Cosmologies in dilatonic Einstein Gauss-Bonnet (dEGB) gravity, where the Gauss–Bonnet term is non–minimally coupled to a scalar field with vanishing potential.
We put constraints on the model parameters when the ensuing modified cosmological scenario drives the WIMP annihilation cross section beyond the present bounds from DM indirect detection searches. In our analysis we assumed WIMPs that annihilate to Standard Model particles through an s-wave process.
 For the class of solutions that comply with WIMP indirect detection bounds, we find that dEGB typically plays a mitigating role on the scalar field dynamics at high temperature, slowing down the speed of its evolution and reducing the enhancement of the Hubble constant compared to its standard value. For such solutions, we observe that the corresponding boundary conditions at high temperature correspond asymptotically to a vanishing deceleration parameter $q$, so that the effect of dEGB is to add an accelerating term that exactly cancels the deceleration predicted by General Relativity.
The bounds from WIMP indirect detection are nicely complementary to late-time constraints from compact binary mergers. This suggests that it could be interesting to use other Early Cosmology processes to probe the dEGB scenario.
}

\begin{document}
\preprint{CQUeST-2023-0720}
\maketitle

\section{Introduction}
\label{sec:introduction}

The difficulty to fit General Relativity (GR) with the other
fundamental interactions of the Standard Model of Particle Physics
(SM) and the failure of the latter to stabilize at the weak scale
(naturalness problem) imply that in spite of their great effectiveness
to describe the observed Universe both theories are believed to be
incomplete. Cosmology and Astrophysics represent excellent testbeds
to probe possible extensions of both GR and the SM.

Indeed, the discovery of Gravitational Waves (GW) and the direct
measurement of merger events of compact binaries has opened up a new
era of precision tests of GR that complement constraints from
Cosmology~\cite{Clifton:2011jh,Nojiri:2010wj} and from laboratory and planetary observations~\cite{Will:2014kxa}. 
In this context, a particularly effective approach to probe
extensions of GR using observational data is the use of effective
models.

Among effective modifications of GR, higher curvature terms are expected to appear in extensions of Einstein Gravity such as string theory and to become important in the early Universe. In particular, Horndeski’s theory is the most general scalar-tensor theory having equations of motion with second-order time derivatives in four-dimensional spacetime~\cite{Horndeski:1974wa}, in which the theory does not have a ghost state~\cite{Woodard:2015zca}. Examples of Horndeski’s theory include additional scalar fields (quintessence~\cite{Tsujikawa:2013fta, Harko:2013gha, Cicoli:2018kdo, Bahamonde:2017ize, Alexander:2019rsc,  Banerjee:2020xcn}, $f(\phi)R$ gravity~\cite{Odintsov:2018ggm}) and/or curvature terms ($f(R)$ gravity~\cite{Nojiri:2006ri}). At the level of the equation of motion~\cite{Horndeski:1974wa, Woodard:2015zca} the simplest example of Horndeski’s theory containing higher--curvature terms is the dilaton-Einstein-Gauss-Bonnet (dEGB) theory, obtained by adding a specific quadratic combination of the curvature non-minimally coupled to a scalar field~\cite{Hwang:1999gf, Satoh:2008ck} (without such non--minimal coupling the Gauss-Bonnet term becomes a topological invariant that does not affect the dynamics). The dEGB theory has been extensively studied in various realizations~\cite{Zwiebach:1985uq,Kanti:1995vq, Hwang:1999gf, Cai:2001dz,
Satoh:2008ck, Guo:2010jr, Koh:2014bka,   Cognola:2006sp,Ahn:2014fwa,Khimphun:2016gsn, Lee:2016yaj, Antoniou:2017acq,Doneva:2017bvd,
Silva:2017uqg,Myung:2018iyq, Lee:2018zym, Chew:2020lkj,Lee:2021uis,Kawai:2021edk,Papageorgiou:2022umj}. In particular, constraints on the dEGB scenario have already been derived using the observed GW signal from Black Hole (BH-BH) or Black Hole-neutron star (BH-NS) merger events~\cite{Nair:2019iur, Okounkova:2020rqw, Wang:2021jfc, Perkins:2021mhb, BH-NS_GB_2022}.

At the temperature $T_{\rm BBN}\simeq$ 1 MeV, Big Bang Nucleosynthesis (BBN) is the earliest process in Cosmology providing a successful confirmation of both GR and the SM, strongly constraining any departure from Standard Cosmology~\cite{Kusakabe:2015ida, Asimakis:2021yct}. On the other hand, although we have no direct probe of the Universe expansion rate, composition or reheating temperature before BBN, an understanding of the present Universe cannot dispense from encompassing Inflation,  Dark Matter (DM) and Baryon asymmetry. All events that take place at $T>T_{\rm BBN}$ can be used to shed light on physics beyond GR and the SM. An explicit example of such approach is, for instance, to identify the dilaton scalar field with the inflaton and to work out observational constraints on dEGB inflation~\cite{Guo:2010jr,Koh:2014bka,Amendola:2005cr}.

Specifically, in the present paper we wish to use the physics of Weakly Interacting Massive Particles (WIMPs) decoupling to probe dEGB Cosmologies (for examples of analyses where the WIMP relic density has been studied within the context of modified Cosmologies see~\cite{Salati:2002md, Rosati:2003yw, Kang:2008zi, Capozziello:2012uv, Capozziello:2015ama, Meehan:2015cna, Lambiase:2016log, profumo_relentless_2017}). WIMPs are the most popular candidates to provide the Cold Dark Matter (CDM) which is supposed to account for about 25\% of the observed energy density of the Universe and explain Galaxy formation. In particular, CDM is one of the most important open issues in modern physics, since the SM provides no candidate for it and as a consequence it requires to introduce exotic physics. WIMPs are expected to have a mass $m_\chi$ in the GeV-TeV range and provide the correct relic abundance $\Omega_\chi h^2 \simeq 0.12$ through thermal decoupling with the relativistic plasma for an annihilation cross section with SM particles $\langle \sigma v\rangle_f\simeq 3\times 10^{-26}$ cm$^3$s$^{-1}$, with the standard assumption that the energy density of the Universe is dominated by radiation and the Hubble constant scales with the temperature $T$ as $H(T) \sim T^2$ at the freeze-out temperature $T_f\simeq m_\chi/20\simeq$ 50 MeV $\sim$ 50 GeV\footnote{As usual, $\Omega_\chi = \rho_\chi/\rho_c$ with $\rho_\chi$ the WIMP number density,  $\rho_c$ = 1.8791 $h^2 \times 10^{-29}$ g cm$^{-3}$ the critical density of the Universe and  $h$ = $H_0$/100 km s$^{-1}$ Mpc$^{-1}$ with $H_0$ the Hubble parameter at present time.}. 

Experimental bounds on such scenarios can be obtained by exploiting the fact that the value of $\langle \sigma v\rangle_f$ that corresponds to the correct CDM relic abundance becomes larger for Cosmologies that enhance $H(T)$ at the time of the WIMP freeze-out, eventually driving the WIMP annihilation rate in our Galaxy today beyond the observational limits on photons, electrons (positrons), (anti)protons and (anti)neutrinos fluxes. For instance, this approach has already been used in the literature to constrain scenarios where $H(T)$ is modified using simple phenomenological parameterizations~\cite{Schelke_2006,Donato:2006}. In the following, we will apply the same strategy to constrain the dEGB scenario. As we will show, due to the scalar field evolution, dEGB presents a high degree of non linearity that implies a more complex phenomenology compared to simplified phenomenological realizations. In particular, this will require to solve the coupled differential equations that drive the WIMP freeze-out process and the scalar field evolution numerically. Interestingly, we will find a degree of complementarity between Late Universe constraints and this specific class of Early Cosmology bounds. This suggests that it could be interesting to further pursue such approach using other Early Cosmology processes, such as for instance thermal leptogenesis, that would allow to probe dEGB at even higher temperatures~\cite{Buchmuller:2005eh,Kawai:2017kqt}.

The paper is organized as follows. In Section~\ref{sec:GB_theory} we outline the dEGB scenario of modified gravity and fix our notations. Section~\ref{Sec:WIMP} is devoted to a summary of the physics that drives the WIMP thermal relic density (Section~\ref{sec:relic}) and the present indirect signals in our Galaxy that can be used to constrain WIMPs (Section~\ref{sec:indirect}). Our quantitative results are contained in Section~\ref{sec:results}. Specifically, the numerical solution of Friedmann equations in dEGB Cosmology is discussed in Section~\ref{sec:numerical}, the Late Universe constraints from BH-BH and BH-NS merger events are summarized for convenience in Section~\ref{sec:BH_NS}, while our main results, i.e. the bounds on the dGEB parameter space from WIMP indirect detection, are discussed in Section~\ref{sec:WIMP_constraints} and combined with those from Section~\ref{sec:BH_NS}. Our Conclusions are provided in Section~\ref{sec:conclusions}, while in Appendix~\ref{app:semi} we provide a qualitative insight on the class of solutions that are obtained numerically by making use of semi-analytical expressions. 

\section{Summary of Gauss-Bonnet theory}
\label{sec:GB_theory}

In this Section we briefly summarize the dEGB scenario and fix our notations (more details can be found in~\cite{Kanti:1995vq,Lee:2016yaj,Koh:2014bka,Myung:2018iyq,Lee:2018zym,Chew:2020lkj,Lee:2021uis}).  The action is given by: 
\begin{equation}
S=  \int_{\mathcal M} \sqrt{-g} \ d^4 x \left[ \frac{R}{2\kappa}
-\frac{1}{2}{\nabla_\mu}\phi {\nabla^\mu}\phi -V(\phi) + \ f(\phi)R^2_{\rm GB} + {\mathcal L}^{\rm rad}_{m} \right]\,,
\label{f-action}
\end{equation} 
 where $\kappa \equiv 8\pi G$ = $1/M_{PL}^2$ (with $M_{PL}$ the reduced Planck mass), $R$ denotes the scalar curvature of the spacetime $\mathcal M$, $R^{2}_{\rm GB}
= R^2 - 4 R_{\mu\nu}R^{\mu\nu} + R_{\mu\nu\rho\sigma} R^{\mu\nu\rho\sigma}$
is the Gauss-Bonnet term and ${\mathcal L}^{\rm rad}_{m}$ describes the interactions of radiation and matter fields.

The coupling between the scalar field and the Gauss-Bonnet term is driven by a function of the scalar field $f(\phi)$. If it is chosen to be a constant, the Gauss-Bonnet term doesn't contribute to the equations of motion being a surface term. The theory can then be reduced to a quintessence model~\cite{Tsujikawa:2013fta, Harko:2013gha, Cicoli:2018kdo, Bahamonde:2017ize, Alexander:2019rsc,  Banerjee:2020xcn}. The coupling function is in principle arbitrary.  An exponential form arises within theories where gravity is coupled to the dilaton~\cite{Kanti:1995vq, Antoniou:2017acq, Lee:2018zym, Lee:2021uis}.  A power law has also been adopted in the literature~\cite{Doneva:2017bvd, Silva:2017uqg} (the two forms are connected by a field redefinition). In the present analysis, we will then adopt the following dEGB realization:
\bea
f(\phi) = \alpha e^{\gamma\phi}.
\label{eq:f_GB}
\eea
\noindent Notice that in string theory the natural sign of the $\alpha$ coefficient is positive~\cite{Boulware:1985wk}. In Refs.~\cite{Koh:2014bka, Lee:2018zym, Lee:2021uis} it was shown that for both signs of $\alpha$ black-hole solutions can be found. In our phenomenological analysis we will adopt both signs.

By varying the action one obtains the equation of motion of the scalar field:
\begin{equation}
\square \phi - V' + f' R^{2}_{\rm GB} =0\,,
\label{s-field}
\end{equation}
\noindent and Einstein's equations:
\begin{equation}
R_{\mu\nu}-\frac{1}{2}g_{\mu\nu}R = \kappa \left( T^{\phi}_{\mu\nu} + T^{\rm GB}_{\mu\nu} + T^{\rm rad}_{\mu\nu} \right) \equiv \kappa T^{\rm tot}_{\mu\nu} \,,
\label{Ei-field}
\end{equation}
where $\square={\nabla_\mu}{\nabla^\mu}$, $V'=\partial V/\partial\phi$ and $f'=\partial f/\partial\phi$. We moved all the additional terms to the right hand side so that it is in the familiar form of the Einstein Equation. 

\noindent The total energy-momentum tensor $T^{\rm tot}_{\mu\nu}$ consists of three terms: $T^{\phi}_{\mu\nu}$ arises from the scalar field action:
\begin{equation}
T^{\phi}_{\mu\nu}=\nabla_{\mu}\phi \nabla_{\nu}\phi - \left( \frac{1}{2}\nabla_{\rho}\phi \nabla^{\rho}\phi + V \right) g_{\mu\nu} \,,
\end{equation}
$T^{\rm GB}_{\mu\nu}$ from the formal contribution of the dEGB term:
\begin{eqnarray}
T^{\rm GB}_{\mu\nu} &=& 4\left[ R \nabla_\mu \nabla_\nu f(\phi) - g_{\mu\nu} R \square f(\phi) \right] -8 \left[
 R_\nu{}^\rho \nabla_\rho \nabla_\mu f(\phi) + R_\mu{}^\rho \nabla_\rho \nabla_\nu f(\phi) \right. \nonumber \\
&& \left. \quad - R_{\mu\nu} \square f(\phi) - g_{\mu\nu} R^{\rho\sigma} \nabla_\rho \nabla_\sigma f(\phi)
+ R_{\mu\rho\nu\sigma} \nabla^\rho \nabla^\sigma f(\phi) \right] \,,
\label{GB-emt}
\end{eqnarray}
\noindent  while $T^{\rm rad}_{\mu\nu}$ is the usual energy-momentum tensor for radiation:
\begin{equation}
T^{\rm rad}_{\mu\nu}=-2\frac{{\delta \cal L}^{\rm rad}_m }{\delta g^{\mu\nu}}+ {\cal L}^{\rm rad}_m g_{\mu\nu}. 
\label{rad-emt}
\end{equation}
\noindent We take the spatially flat Friedmann-Lema\^{i}tre-Robertson-Walker (FLRW) metric:
\begin{equation}
ds^2 = - dt^2 + a^2(t)\delta_{ij} dx^i dx^j\,, \label{frwmewtric}
\end{equation}
\noindent that implies that the scalar field depends only on time, $\phi=\phi(t)$. The energy density and the pressure can be obtained from the energy-momentum tensor as $-\rho_{I}={{T_{I}}^0}_0 $ and  ${{T_{I}}^i}_j = p_{I}{\delta^i}_j$, where $I=\{\phi, \, {\rm GB}, \, {\rm rad}\}$. Specifically, the energy density and pressure for the scalar are
\begin{equation}
\rho_{\phi}=\frac{1}{2}\dot{\phi}^2+V(\phi),  {\quad} p_{\phi}=\frac{1}{2}\dot{\phi}^2-V(\phi),
\label{rhopi_phi}
\end{equation}
\noindent while the corresponding formal quantities from the Gauss-Bonnet term are:
\begin{eqnarray}
\rho_{\rm GB} &=& -24\dot{f} H^3 = -24f^\prime{\dot{\phi}} H^3=-24{\alpha}\gamma e^{\gamma\phi}\dot{\phi}H^3.
\label{eq:rho_GB}
\\
p_{\rm GB} &=& 8 \left(f^{\prime\prime}\dot{\phi}^2+f^\prime\ddot{\phi}\right) H^2 + 16 {f^\prime}\dot{\phi} H(\dot{H}+H^2)\nonumber\\
&=& 8\frac{d(\dot{f}H^2)}{dt}+16 \dot{f}H^3 =8\frac{d(\dot{f}H^2)}{dt} -\frac{2}{3} \rho_{\rm GB}\,.
\label{eq:p_GB}
\end{eqnarray}
\noindent Here, $H(t) = \frac{{\dot a}(t)}{a(t)}$ is the Hubble parameter, $a$ the scale factor, $ {\dot f}=f' {\dot \phi}$ and $ {\ddot f}=f''{\dot \phi}^2+f' {\ddot \phi}$. The radiation energy density $\rho_{\rm rad}$ and pressure $p_{\rm rad}$ from Eq.~(\ref{rad-emt}) are taking contribution from all relativistic species satisfying the equation of state $p_{\rm rad}=\frac{1}{3} \rho_{\rm rad}$.
The energy density $\rho_{\rm rad}$ at temperature $T$ is given by $\rho_{\rm rad} = \frac{\pi^2}{30}\,g_{*}\,T^4$ with $g_{*}$ the number of effective relativistic degrees of freedom in equilibrium with the thermal bath. 
The total energy-momentum tensor satisfies the continuity equation. Specifically, radiation satisfies: 
\begin{equation}
\dot{\rho}_{\rm rad} + 3 H (\rho_{\rm rad} + p_{\rm rad}) = 0. 
\end{equation}
Similarly, the sum of the scalar and the Gauss-Bonnet contributions can be shown to satisfy: 
\begin{equation}
\dot{\rho}_{\{\phi + \rm GB\}} + 3 H \left( {\rho}_{\{\phi + \rm GB\}} + {p}_{\{\phi + \rm GB\}} \right)= 0,
\label{conti_phiGB}
\end{equation}
where the subscript $\{{\phi + \rm GB}\}$ represents the sum of the contributions from the scalar field and the Gauss-Bonnet term, ${\rho}_{\{\phi + \rm GB\}} \equiv {\rho}_{\phi} +{\rho}_{\rm GB}$ and ${p}_{\{\phi + \rm GB\}} \equiv {p}_{\phi} +{p}_{\rm GB}$.
However, the scalar and Gauss-Bonnet contributions don't satisfy the continuity equation separately. This is due to the interaction between the Gauss-Bonnet term and the scalar field.  It is also worthwhile to mention that the signature of ${\rho}_{\{\phi + \rm GB\}}$ and ${p}_{\{\phi + \rm GB\}}$ is not necessarily positive.

\noindent The Friedmann equations can be written as~\cite{Koh:2014bka}:
\begin{eqnarray}
H^2 &=& \frac{\kappa}{3} \left( {\rho}_{\{\phi + \rm GB\}} +\rho_{\rm rad}\right) \equiv  \frac{\kappa}{3} \rho_{\rm tot}\,,
\label{Eqfrw1}
\\
{\dot H} &=& - \frac{\kappa}{2} \left[ ({\rho}_{\{\phi + \rm GB\}} + {p}_{\{\phi + \rm GB\}}) +(\rho_{\rm rad} + p_{\rm rad}) \right] \equiv - \frac{\kappa}{2} (\rho_{\rm tot}+p_{\rm tot}) \,,
\label{Eqfrw2} \\
{\ddot \phi} &+& 3H {\dot \phi} + V' -f' R^{2}_{\rm GB} =0\,,
\label{Eqfrw3}
\end{eqnarray}
\noindent where $\rho_{\rm tot}$ and $p_{\rm tot}$ can be interpreted as the total energy density and the pressure of the Universe. 
The combination of Eqs.~(\ref{Eqfrw1}) and (\ref{Eqfrw2}) gives the acceleration (deceleration) of expansion  
\begin{eqnarray}
{\dot H}+H^2 &=& \frac{\ddot{a}}{a} \equiv - H^2 q \nonumber \\
    &=& -\frac{\kappa}{6} \left( ({\rho}_{\{\phi + \rm GB\}} + 3{p}_{\{\phi + \rm GB\}}) +(\rho_{\rm rad} + 3p_{\rm rad}) \right) \nonumber\\
    &=& -\frac{\kappa}{6} \rho_{\rm tot} (1+3w_{\rm tot}) = -\frac{1}{2} H^2 (1+3w_{\rm tot}), 
    \end{eqnarray}
    
\noindent where:
\begin{equation}
    q=-\frac{\ddot{a}a}{\dot{a}^2} =\frac{1}{2} (1+3w_{\rm tot}), 
    \label{decel_par}
\end{equation}
   
\noindent is the deceleration parameter and $w_{\rm tot}=\frac{p_{\rm tot}}{\rho_{\rm tot}}$ is 
the equation of state of the Universe. The explicit form of the energy densities and pressures are already given in Eqs.~(\ref{rhopi_phi}), (\ref{eq:rho_GB}) and (\ref{eq:p_GB}). The Gauss-Bonnet term is written as $R^{2}_{\rm GB}= 24 H^2({\dot H} + H^2)\equiv -24 H^4 q$.
In particular, it is the scalar field equation~(\ref{Eqfrw3}) that makes the continuity equation~(\ref{conti_phiGB}) for the sum of the scalar and the Gauss-Bonnet term hold. This can be seen by observing that
$ \dot{\rho}_{\phi}  + 3 H ( {\rho}_{\phi} + {p}_{\phi} )= {\dot \phi}({\ddot \phi} + 3H {\dot \phi} + V') $ 
and
$\dot{\rho}_{\rm GB} + 3H ({\rho}_{\rm GB}+{p}_{\rm GB}) =  
 -24 \dot{\phi}f' H^2({\dot H} + H^2) =-\dot{\phi}f' R^{2}_{\rm GB}.$ 

Then the Friedmann equations take the explicit form:
\begin{eqnarray}
H^2 &=& \frac{\kappa}{3} \left( \frac{1}{2} {\dot\phi}^2 +V -24 {\dot f} H^3  +\rho_{\rm rad}\right) \,,
\label{eqfrw1}
\\
{\dot H} &=& - \frac{\kappa}{2} \left( {\dot \phi}^2 +8\frac{d(\dot{f}H^2)}{dt}-8\dot{f}H^3 +\rho_{\rm rad} + p_{\rm rad} \right) \,,
\label{eqfrw2} \\
{\ddot \phi} &+& 3H {\dot \phi} + V' -24 f' H^2({\dot H} + H^2)=0\,.
\label{eqfrw3}
\end{eqnarray}

\noindent Note that Eq.~(\ref{Eqfrw2}) or Eq.~(\ref{eqfrw2}) can be obtained by taking a time derivative of Eq.~(\ref{Eqfrw1}) or Eq.~(\ref{eqfrw1}), 
and using Eq.~(\ref{eqfrw3}) and the continuity equation for radiation.

In Eq.~(\ref{f-action}) the potential $V(\phi)$ is in principle an arbitrary function of $\phi$ for which a power law or an exponential form is adopted in the literature and that, for instance, plays a crucial role in inflationary scenarios. In our case it represents a non-minimal ingredient that enlarges the number of free parameters without being strictly necessary, while interfering with the peculiar non-linear effect of the GB term on the evolution of the scalar field $\phi$ that represents the main characteristic of the dEGB scenario that we wish to discuss. Moreover, in order to avoid early accelerated expansion before matter-radiation equivalence it is crucial that $V$ is exactly vanishing or extremely close to zero at $T$ = $T_{\rm BBN}$, requiring to either tune the scalar field evolution or to adopt an ad-hoc functional form that arbitrarily sets $V$ to zero below some threshold temperature. For these reasons with the goal of naturalness and minimality we choose to assume $V=0$ in our analysis.  

\section{WIMP thermal relic density and indirect detection bounds}
\label{Sec:WIMP}

\subsection{WIMP relic density}
\label{sec:relic}

In the thermal decoupling scenario, the WIMP number density $n_\chi$ closely follows its equilibrium value $n_{eq}$ as long as $T/m_\chi \gtrsim$ 1, because the rate of WIMP annihilations to SM particles $\Gamma=n_\chi\langle \sigma v\rangle$ is larger than the expansion rate of the Universe $H$. When $T/m_\chi \lesssim$ 1 the equilibrium density $n_{eq}$ becomes exponentially suppressed and eventually, at the freeze-out temperature $T$ = $T_f\simeq m_\chi/20$, the ratio $\Gamma/H$ drops below 1 and the WIMPs stop annihilating so that their number in a comoving volume remains approximately constant. In the standard scenario, the WIMP freezes out in a radiation background, i.e. when  $H\equiv H_{\rm rad}\sim T^2/\Mp$ and the correct prediction for the observed CDM density $\Omega h^2 \simeq 0.12$ is obtained for $\langle \sigma v\rangle_f\simeq 3\times 10^{-26}$ cm$^3$s$^{-1}$. In particular, for a larger value of $\langle \sigma v\rangle_f$ the WIMPs number density $n_\chi$ follows its exponentially suppressed equilibrium value $n_{eq}$ for a longer time, leading to a smaller comoving density at decoupling. This implies that the relic density is anti-correlated to the annihilation rate, i.e. $\Omega h^2\sim 1/\langle \sigma v\rangle_f$. Expanding $\langle \sigma v\rangle$ in powers of $v^2/c^2\ll$1 one gets~\cite{Choi:2017mkk}:
\begin{equation}
\langle{\sigma v}\rangle \simeq a + 6\,b\, \frac{T}{m_{\chi}},
\label{eq:s_p_wave}
\end{equation}
\noindent where the contributions from s-wave ($l = 0$) and p-wave ($l = 1$) annihilations are determined by the two constants $a$ and $b$. In a standard radiation background, WIMPs completely stop annihilating after freeze-out because the annihilation rate redshifts faster than the Hubble rate. In particular, using $H\sim T^2$ and $n_\chi\sim a^{-3}\sim T^3$ one gets $\Gamma/H_{\rm rad}$  = $n_\chi \langle{\sigma v}\rangle/H_{\rm rad}\sim T$   for s-wave ($a\ne$ 0) and $\Gamma/H_{\rm rad}\sim T^{2}$ for p-wave ($a$ = 0).  

A modified cosmological scenario affects the relic density if it changes the Universe expansion rate at the WIMP freeze-out temperature compared to the standard radiation background. In particular, if the WIMP decouples when $H(T^\prime)\sim H(T) (T^{\prime}/T)^\xi$ with $\xi>2$ the freeze-out condition $\Gamma/H=1$ is achieved at a larger temperature, so that at fixed $\langle \sigma v\rangle_f$ the relic density is increased. This implies that the correct relic density is achieved for a larger value of $\langle \sigma v\rangle_f$. A more subtle effect~\cite{profumo_relentless_2017} is that in this case for $T<T_f$ one has $\Gamma/H\sim T^{3-\xi}$ (s-wave) or $\Gamma/H\sim T^{4-\xi}$ (p-wave) so that if $\xi$ is large enough  ($\xi>3$ for s-wave and $\xi>4$ for p-wave) the post freeze-out annihilation rate redshifts slower than the Hubble rate. When this happens, at variance with the case when they decouple in a radiation background, the WIMP particles keep annihilating also after freeze-out, substantially reducing the final value of the relic density. This reduction partially mitigates the enhancement effect due to the anticipated freeze-out.

The Boltzmann equation describing the evolution of the WIMP number density is conveniently expressed in terms of the comoving density $Y\equiv n_\chi/s$, with $s = (2 \pi^2 / 45) g_{*s} T^3$ being the entropy density of the Universe and $g_{*s}(T)$ the corresponding number of degrees of freedom, equal to $g_{*}$ before neutrino decoupling~\cite{g_T_Steigman2012}. Indicating with $Y_{eq}$ the corresponding equilibrium value we can write:
\begin{eqnarray}
\dfrac{dY_{\chi}}{dx} = -\dfrac{\beta\,s}{H\,x}{\langle{\sigma\,v}\rangle}
\left(Y_{\chi}^2 - (Y^{\rm eq}_{\chi})^2\right)\,,
\label{eq:BE}
\end{eqnarray}
\noindent where $x=m_{\chi}/T$ and:
\begin{eqnarray}
\beta = \left(1+\dfrac{1}{3}\dfrac{d \ln g_s}{d \ln T}\right)\,.
\end{eqnarray}

The temperature evolution of the Hubble parameter is obtained by solving the coupled differential
equations for $\dot{H}$ and $\ddot{\phi}$ (Eqs.\,\,(\ref{eqfrw2}) and (\ref{eqfrw3}), respectively). Assuming iso-entropic expansion ($sa^3$ = constant), the change of variable from time to temperature is given as usual as:
\begin{eqnarray}
\dfrac{dT}{dt} = -\dfrac{H\,T}{\beta}\,.
\label{eq:T-t}
\end{eqnarray}

In Section~\ref{sec:results}  we obtain the WIMP comoving density $Y^0_{\chi}$ at present time 
in terms of $\langle \sigma v\rangle$ and $m_\chi$ by solving Eq.~(\ref{eq:BE}) numerically. Finally, in terms of $Y^0_{\chi}$ the WIMP relic density is given by:
\begin{eqnarray}
\Omega_{\chi} h^2 = \dfrac{\rho_{\chi}}{\rho_c} h^2 = 2.755 \times
\left(\dfrac{m_{\chi}}{\rm GeV}\right) Y^0_{\chi}\,.
\end{eqnarray}
\noindent In particular we will indicate with $\langle \sigma v\rangle_{\rm relic}$ the value of $\langle \sigma v\rangle_{f}$ that yields the observed value of the present DM relic density, $\Omega_{\chi} h^2 = 0.12$. 

\subsection{Indirect detection bounds on WIMP annihilation}
\label{sec:indirect}
The same annihilation processes to SM particles that keep WIMPs in thermal equilibrium in the Early Universe can be used to search for indirect detection signals at later times.  

WIMPs annihilations in the halo of our Galaxy can produce various primary SM particles which can further cascade leading to secondary final states such as photons, electrons (positrons), (anti)protons and (anti)neutrinos. Such particles can be looked for using different types of experiments.
For example, positrons or antiprotons with energies above a few tens of GeV 
can be searched for using cosmic-ray spectrometers like AMS~\cite{AMS_2014, AMS_2019, AMS_DM_2017}. On the other hand, gamma-ray telescopes like 
Fermi LAT~\cite{Fermi-LAT_2017GC, Fermi-LAT_2017} can look for gamma rays from WIMP annihilations in the Galactic center or in dwarf spheroidal galaxies (dSphs). 
Ionizing particles injected by WIMP annihilations can also affect the anisotropies of the Cosmic Microwave Background radiation (CMB) and can be constrained by Planck data~\cite{CMB_2015}. 

Assuming a self-conjugate WIMP the amount of $e^+$ or $\gamma$-rays with energy $E$ produced by WIMP annihilations per unit time,  volume and energy is given by: 
\begin{equation}
Q_{e^+/\gamma}(r,E) = {\langle \sigma v \rangle}_{\rm gal} \frac{\rho^2_{\chi}(r)}{2 m^2_{\chi}} 
\sum_{F} B_F \frac{dN^F_{e^+/\gamma}}{dE} (E,m_{\chi}), 
\label{eq:DM_source_fn}
\end{equation}
where $\langle\sigma v\rangle_{\rm gal}$ is the WIMP annihilation cross section in our Galaxy, $\rho_{\chi}/m_\chi$ is the number density of DM at the location $r$ of the annihilation process,
$B_F$ is the branching fraction to the primary annihilation channel $F$, while ${dN^F_{e^+/\gamma}}/{dE}$ is the $e^+/\gamma$ energetic spectrum per $F$ annihilation.
Depending on $m_{\chi}$ the WIMP primary annihilation channels can be $e^+e^-$, $\mu^+\mu^-$, $\tau^+\tau^-$, $b\bar{b}$,  $t\bar{t}$, $\gamma\gamma$, $W^+W^-$, $ZZ$, etc. Gamma rays propagate on a straight line so their signal points back to the source, 
while the propagation of positrons is bent by the galactic magnetic field and is affected by solar modulation and energy-loss processes along the way~\cite{pppc_2010, Fermi-LAT_2017, indirect_search_Beacom_2018}.

The non-observation of a significant excess over the background allows to use Eq.~(\ref{eq:DM_source_fn}) to obtain an upper bound on $\langle \sigma v\rangle_{\rm gal}$ as a function of $m_\chi$. In particular, in the analysis of Section~\ref{sec:results} we will assume the case of s-wave annihilation ($a\ne 0$ in Eq.~(\ref{eq:s_p_wave})) for which $\langle \sigma v\rangle_{\rm gal}$ = $\langle \sigma v\rangle_f$, with  $\langle \sigma v\rangle_f$ the same annihilation cross section times velocity at freeze-out discussed in Section~\ref{sec:relic}. 

\begin{table*}[ht!]
\begin{center}
\def\arraystretch{1.2}
\begin{tabular}{|c|c|}
\hline 
$m_{\chi}$  [GeV]      & ${\langle \sigma v \rangle}_{\rm ID}$ [$\mbox{cm}^3 \mbox{s}^{-1}$]         \\ 
\hline
10         & $1.8 \times 10^{-26}$                 \\
\hline
100        & $10^{-25}$                            \\     
\hline
1000       & $3 \times 10^{-24}$                   \\
\hline
\end{tabular}
\caption{95\% C.L. conservative indirect detection upper-bound ${\langle \sigma v \rangle}_{\rm ID}$ on ${\langle \sigma v \rangle}_{\rm gal}$ from the analysis of Ref.~\cite{indirect_search_Beacom_2018} for the three WIMP mass benchmarks discussed in Section~\ref{sec:results}.}
\label{tab:mx_sv_ID}
\end{center}
\end{table*}

The bound on $\langle \sigma v\rangle_{\rm gal}$ depends on the branching fractions $B_F$, which for a generic WIMP are not fixed. In this case, a conservative bound on  $\langle \sigma v\rangle_{\rm gal}$ has been obtained in~\cite{indirect_search_Beacom_2018} by scanning over all the possible $0\le B_F\le1$ combinations (with the exception of $F$ = neutrinos) with the constraint $\sum_{F} B_F = 1$. The authors considered an exhaustive list of existing experiments and took conservative assumptions on the astrophysical backgrounds as well as for various quantities such as the DM 
density profile, the Galactic magnetic field and diffusion parameter, solar modulation, etc. 
The resulting 95\% C.L. combined limit on ${\langle \sigma v \rangle}_{\rm gal}$ 
is provided in Fig. 5 of \cite{indirect_search_Beacom_2018}. In Table~\ref{tab:mx_sv_ID} we report from Ref.~\cite{indirect_search_Beacom_2018} the numerical values corresponding to the three benchmark WIMP masses $m_\chi$ = 10 GeV, 100 GeV and 1 TeV that will be used in Section~\ref{sec:results} to constrain the allowed parameter space of the dEGB model.  In this range of $m_\chi$ the bound is mainly determined by the measurement of the positron flux by AMS-02~\cite{AMS_2014, AMS_2019}  and that of the gamma-ray flux from dSphs by Fermi LAT~\cite{Fermi-LAT_2017}.

\section{Results}
\label{sec:results}

\subsection{Numerical solutions of Friedmann equations}
\label{sec:numerical}

In this Section, we discuss the numerical solutions of the modified Friedmann equations presented 
in Section~\ref{sec:GB_theory}. As already pointed out, in our analysis we will assume 
$V (\phi) = 0$ and the dEGB function $f(\phi)$ defined in Eq.~(\ref{eq:f_GB}). As a consequence, the WIMP decoupling mechanism is expected to be modified compared to the standard case. In particular, introducing the enhancement parameter:

\begin{equation}
    A(T)\equiv \frac{H(T)}{H_{\rm rad}},
    \label{eq:enhancement}
\end{equation}

\noindent if $\rho_{\rm tot}> \rho_{\rm rad}$ through Eq.~(\ref{eqfrw1}) one has $A>1$ and the $\langle \sigma v\rangle_f$ = $\langle \sigma v\rangle_{\rm relic}$ corresponding to the observed relic density is driven to higher values compared to the standard case. This will be used in Section~\ref{sec:WIMP_constraints} to constrain the dEGB parameter space using the bounds from WIMP indirect searches.  The results of this Section rely on numerical solutions. In Appendix~\ref{app:semi} a qualitative insight on such solutions is provided making use of some semi-analytical approximations. 

We rewrite the Friedmann equations in Eqs.~(\ref{eqfrw1}), (\ref{eqfrw2}) and (\ref{eqfrw3}) for reference:
\begin{eqnarray}
H^2 &=& \frac{\kappa}{3} \left( -24 f^{\prime}{\dot \phi} H^3 +\frac{1}{2} {\dot\phi}^2   +\rho_{\rm rad}\right) \,, \label{Eq_frw1} \\
{\dot H} &=& - \frac{\kappa}{2} \left( {\dot \phi}^2 +8\frac{d(f^{\prime}{\dot \phi}H^2)}{dt}-8f^{\prime}{\dot \phi}H^3 +\rho_{\rm rad} + p_{\rm rad} \right) \,,
\label{Eq_frw2} \\
{\ddot \phi} &+& 3H {\dot \phi} + V^{\prime}_{\rm GB}=0\,,
\label{Eq_frw3}
\end{eqnarray}
where:
\begin{eqnarray}
V^{\prime}_{\rm GB}&\equiv& -f' R^{2}_{\rm GB}= -24 f' H^2({\dot H} + H^2)=24 {\alpha}\gamma e^{\gamma\phi} q H^4, \,  
\end{eqnarray}
\noindent is technically not the gradient of a potential, but just an excess of notation to indicate that it drives the scalar field evolution, while $q=(1+3w)/2$ is the usual deceleration parameter in (\ref{decel_par}) ($q=1$ for radiation dominance at $T_{\rm BBN}$).
We present our results in geometric units (defined as $\kappa = 8 \pi G = 1$, $c = 1$) in which $\alpha$ is in $\rm m^2$ and $\phi$ as well as $\gamma$ are dimensionless.

Equations (\ref{Eq_frw2}) and (\ref{Eq_frw3}) can be re-arranged into a set of three first order coupled differential equations for the quantities $\phi$, $\dot{\phi}$ and $H$. We set the boundary conditions on $\phi(T_{\rm BBN})$  $\equiv \phi_{\rm BBN}$,  $\dot{\phi}(T_{\rm BBN})$  $\equiv \dot{\phi}_{\rm BBN}$ and $H(T_{\rm BBN})$ $\equiv H_{\rm BBN}$ at the BBN temperature, assumed to be $T = T_{\rm BBN}$ = 1 MeV.
Note that a shift of $\phi_{\rm BBN}$ is equivalent to a redefinition of the $\alpha$ parameter thanks to the fact that $V(\phi) = 0$ and the scalar field $\phi$ (and hence $\phi_{\rm BBN}$) appears in the Friedmann equations (\ref{Eq_frw1})-(\ref{Eq_frw3}) 
only through $f(\phi)$:
\begin{equation}
{\phi^{\prime}}_{\rm BBN} = \phi_{\rm BBN} + \phi_{0},\,\,\,\alpha^{\prime} = \alpha \hspace{0.5mm} e^{-\gamma \phi_{0}},\,\,\,\gamma^{\prime} = \gamma . 
\label{eq:Gauge}
\end{equation}
\noindent In other words, any choice of $\phi_{\rm BBN}$ corresponds to a
specific gauge fixing and the quantity:
\begin{equation}
\tilde{\alpha}=\alpha \hspace{0.5mm} e^{\gamma \phi_{\rm BBN}},
\label{eq:alpha_tilde}
\end{equation}
\noindent is invariant under the gauge transformation~(\ref{eq:Gauge}).
In our analysis, we will show our results in terms of $\tilde{\alpha}$ (which is equivalent to adopt the gauge $\phi_{\rm BBN}= 0$). Notice that the boundary condition on $H(T_{\rm BBN})$ = $H_{\rm BBN}$ can be obtained from $\phi_{\rm BBN}(=0)$ and $\dot{\phi}_{\rm BBN}$ by solving the cubic equation Eq. (\ref{Eq_frw1}) at $T_{\rm BBN}$:
\begin{eqnarray}
&&A H_{\rm BBN}^3+H_{\rm BBN}^2 -B=0,\label{eq:cubic_H} \nonumber \\ 
&& A=8 {\kappa} f^{\prime}(\phi_{\rm BBN}) \dot{\phi}_{\rm BBN}, \\
&& B= \frac{\kappa}{3} \left [\frac{1}{2} \dot{\phi}_{\rm BBN}^2+\rho_{\rm rad}(T_{\rm BBN})\right], \nonumber
\end{eqnarray}
\noindent and taking the positive solution closer to $H_{\rm rad}(T_{\rm BBN})$\footnote{Notice that a cubic equation as always at least one real solution. We find more convenient to include the expression of $\dot{H}$ to the set of differential equations. An alternative way to obtain the evolution of $H(T)$ is to solve the cubic equation (\ref{Eq_frw1}), which is valid at all temperatures, in terms of $\phi(T)$ and $\dot{\phi}(T)$.}. The only boundary condition taking arbitrary value is then $\dot{\phi}_{\rm BBN}$, which can be chosen to be positive (or zero). This is because the configurations with $\dot{\phi}_{\rm BBN}<0$ are obtained when $\gamma\rightarrow -\gamma$ since the solutions of the Friedmann equations become invariant under a simultaneous change of sign of $\dot{\phi}_{\rm BBN}$ and $\gamma$. 

The contribution of $\rhoBBNT=\frac{1}{2}\dot{\phi}_{\rm BBN}^2$ to the energy density at BBN is constrained by the upper bound on the effective number of neutrino flavors $N_{eff}\le$ 2.99 $\pm$ 0.17~\cite{planck_2018}, that can be converted into $\rhoBBNT \le 3 \times 10^{-2} \rho_{\rm BBN}$, with $\rho_{\rm BBN}$ the standard radiation energy density at BBN. In Section~\ref{sec:WIMP_constraints}, we will adopt three nonnegative benchmarks for $\dot{\phi}_{\rm BBN}$ corresponding to $\rhoBBNT =0 $,
the upper bound $\rhoBBNT =3\times 10^{-2}\rho_{\rm BBN}$ and an illustrative intermediate value $\rhoBBNT = 10^{-4}\rho_{\rm BBN}$.  

Once the boundary condition for $\dot{\phi}_{\rm BBN}$ is given for one of the three benchmarks above with the gauge $\phi_{\rm BBN}=0$ and $H_{\rm BBN}$ from Eq.~(\ref{eq:cubic_H}), the solutions of the set of the three first order coupled differential equations for $\phi$, $\dot{\phi}$ and $H$ are obtained for given values of the parameters $\tilde{\alpha}$ and $\gamma$ by evolving them numerically from $T_{\rm BBN} = 1$ MeV to higher temperatures (we stop at $T$ = 100 TeV).
The temperature evolution can be obtained by changing variable from $t$ to $T$ using Eq.~(\ref{eq:T-t}).

In the remaining part of this Section we illustrate numerical solutions and show with some examples how Cosmology is modified at $T>T_{\rm BBN}$ by the dEGB scenario.

\begin{figure}[hbt!]
\centering
\includegraphics[width=7.49cm,height=5cm]{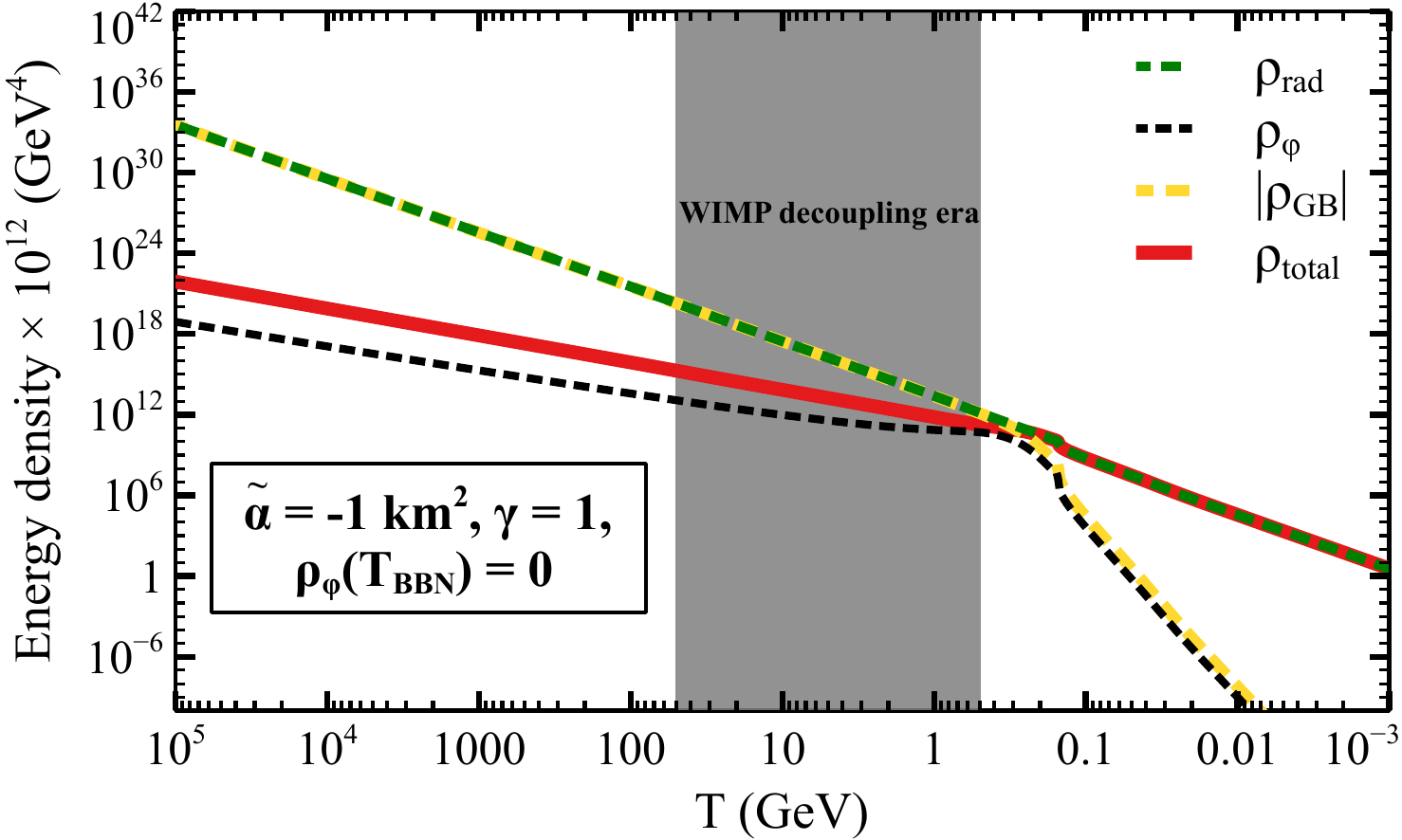}
\includegraphics[width=7.49cm,height=5cm]{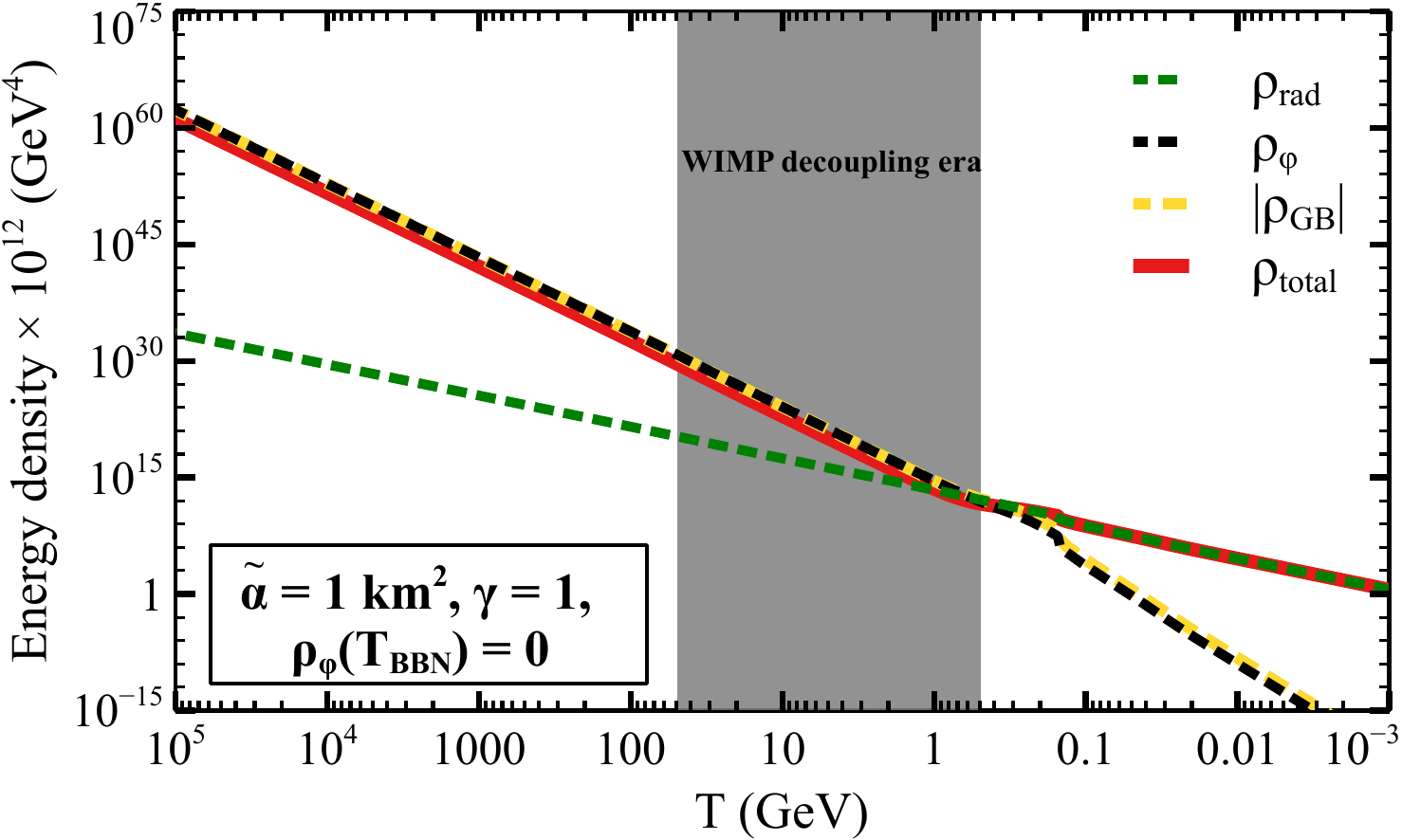}
\caption{Evolution of $\rho_{\rm rad}$, $\rho_{\phi}$, $\rho_{\rm GB}$
and $\rho_{\rm total}$ with $T$ for $\rho_{\phi}(T_{\rm BBN}) = 0$
and $\phi(T_{\rm BBN}) = 0$. }
\label{fig:rho_phidot_0}
\end{figure}

\begin{figure}[hbt!]
\centering
\includegraphics[width=7.49cm,height=5cm]{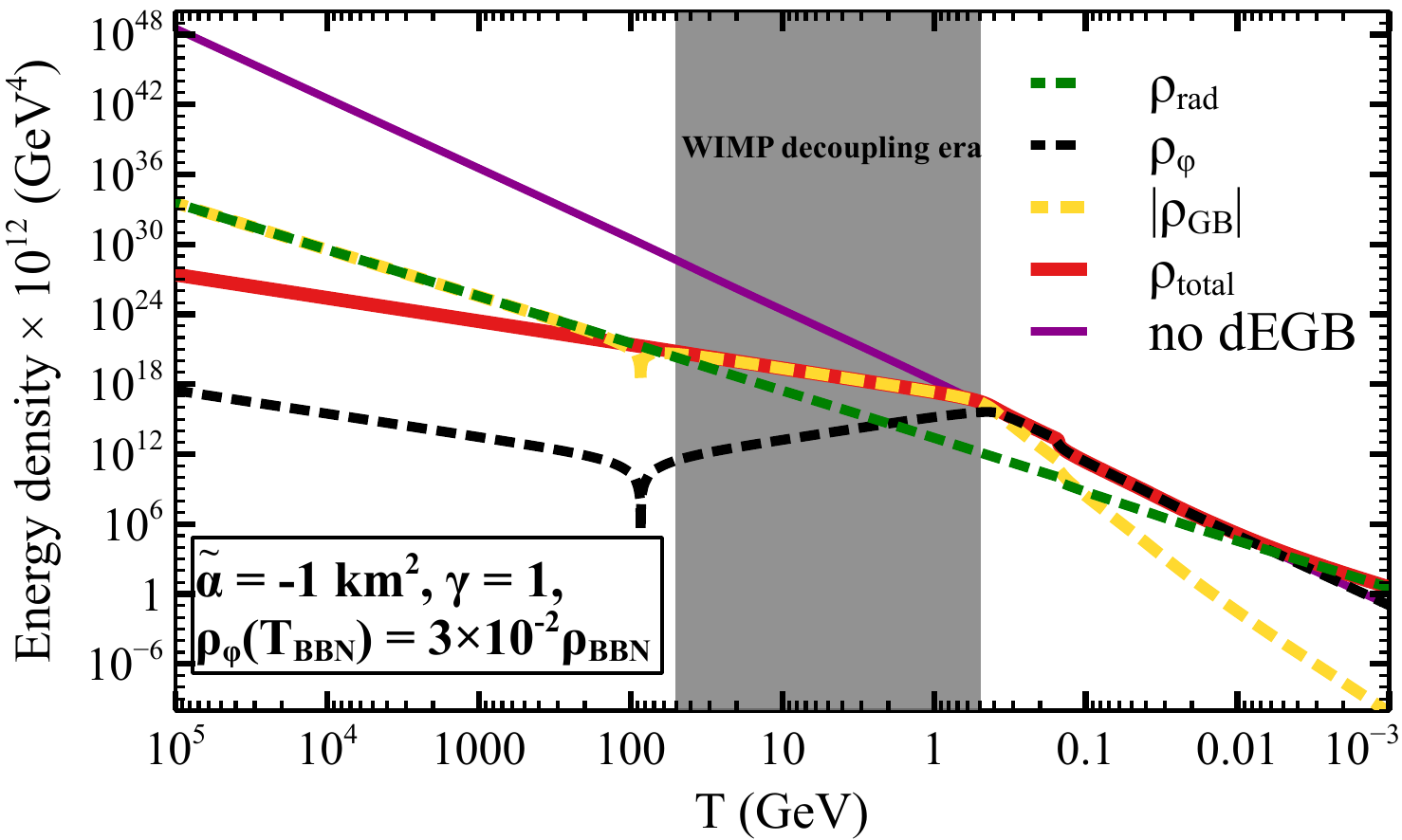}
\includegraphics[width=7.49cm,height=5cm]{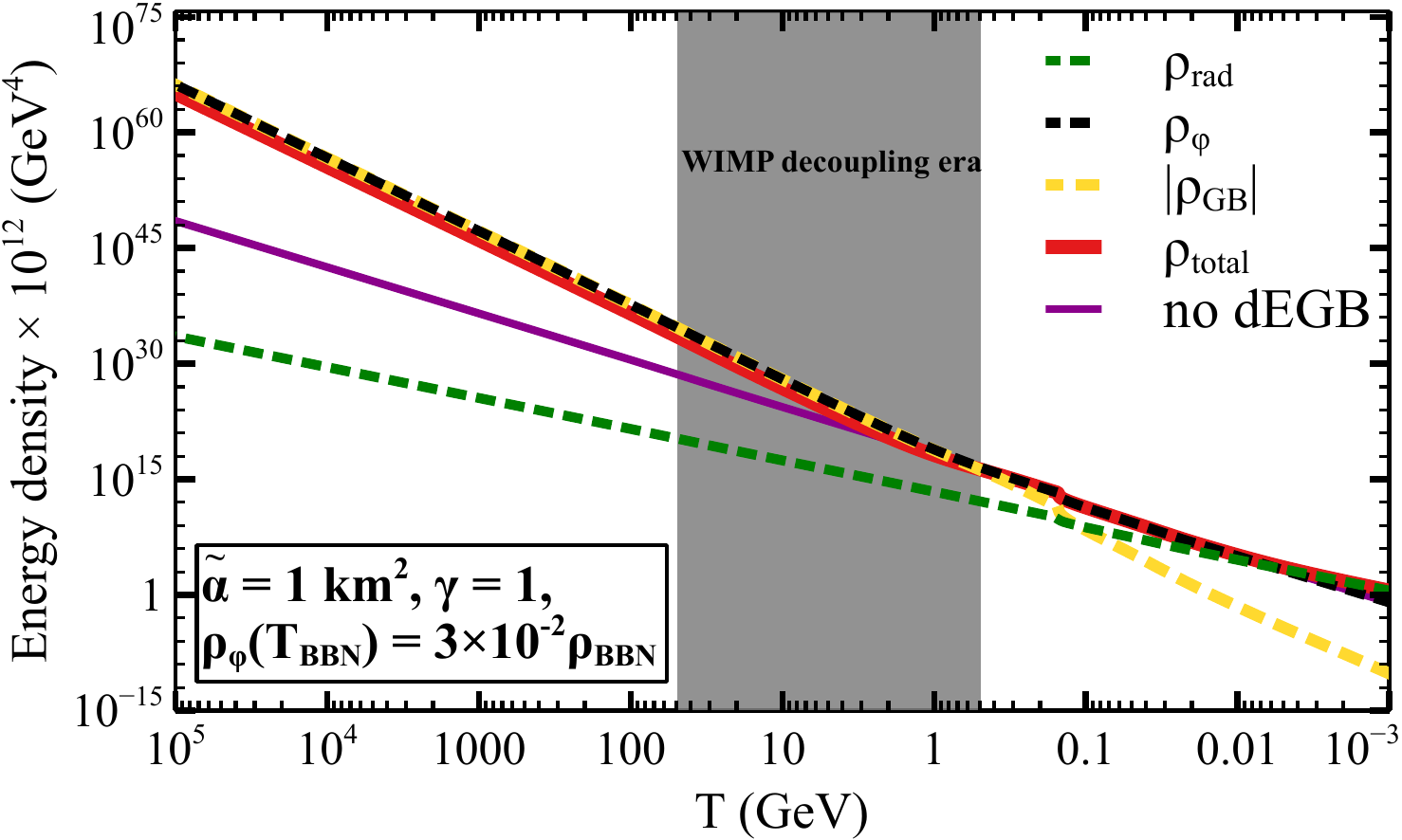}
\includegraphics[width=7.49cm,height=5cm]{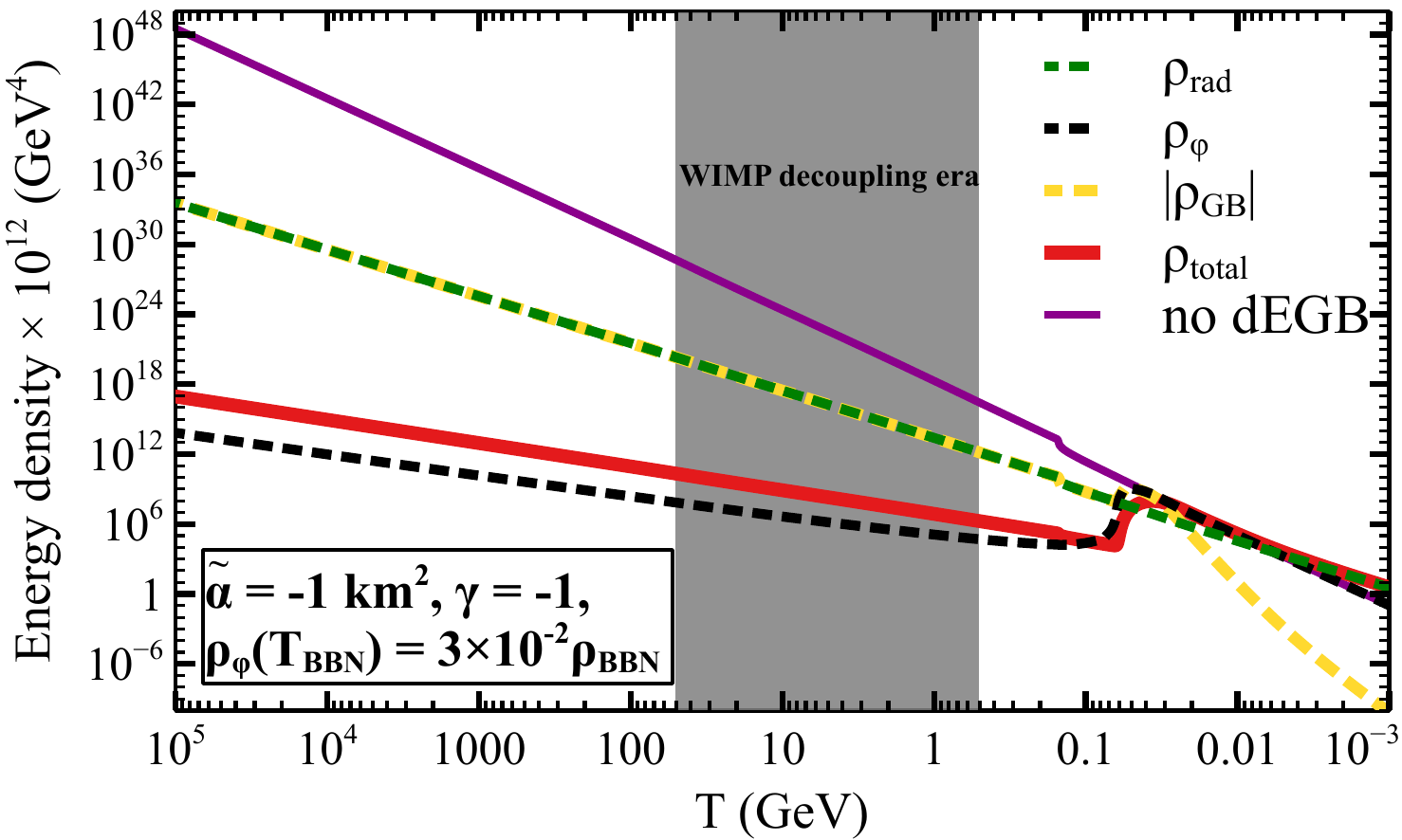}
\includegraphics[width=7.49cm,height=5cm]{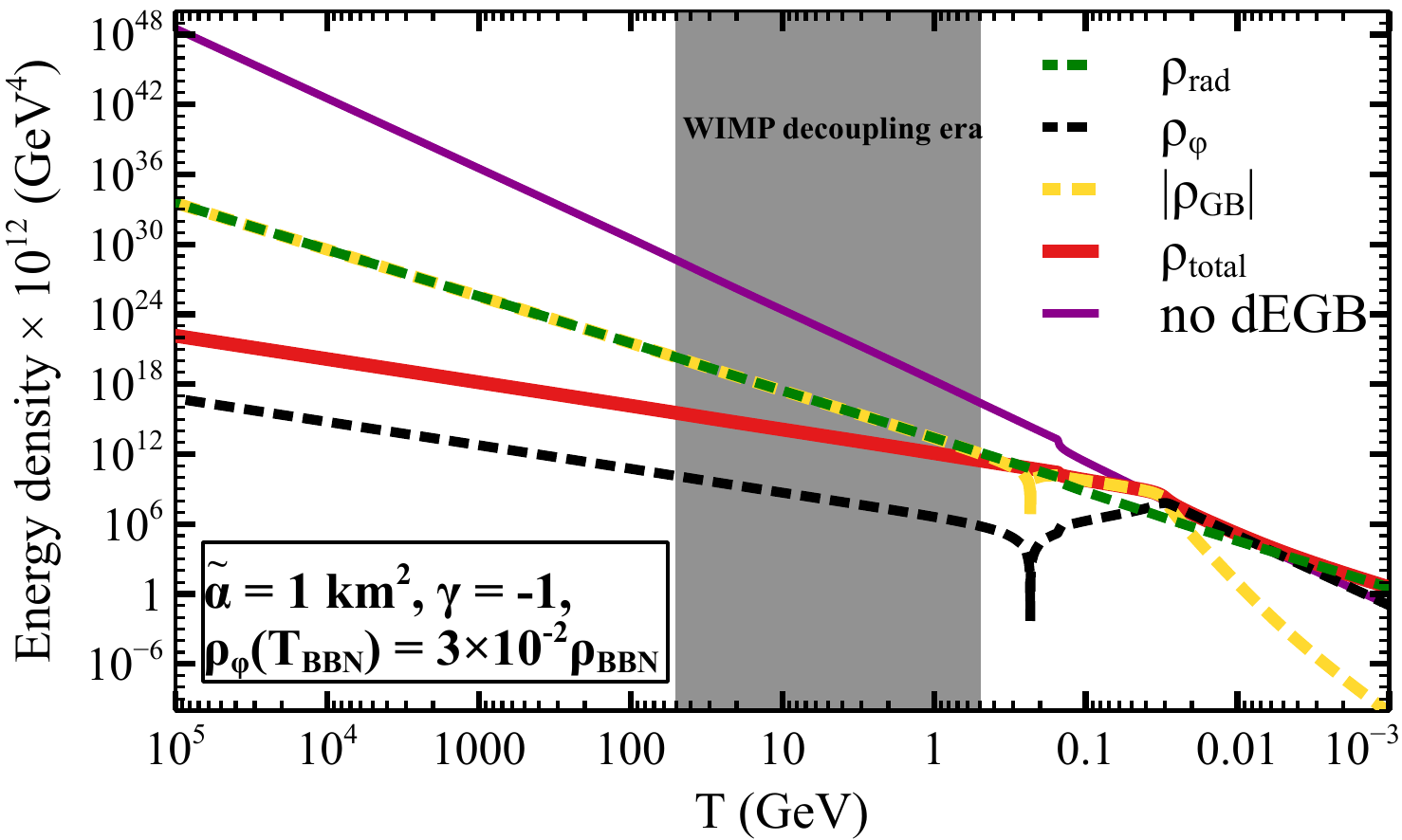}
\caption{Evolution of $\rho_{\rm rad}$, $\rho_{\phi}$, $\rho_{\rm GB}$
and $\rho_{\rm total}$ with $T$ for $\rho_{\phi}(T_{\rm BBN}) = 3\times10^{-2}\rho_{\rm BBN}$
and $\phi(T_{\rm BBN}) = 0$. Recall that only $\rho_{\rm rad}$ and $\rho_{\rm total}$ are physical quantities.}
\label{fig:rho_phidot_neq0}
\end{figure}

In Figs.~\ref{fig:rho_phidot_0} and \ref{fig:rho_phidot_neq0} we show the evolution of the energy density of the Universe $\rho_{tot}$ and of its different contributions, as indicated in Eq.~(\ref{Eq_frw1}), for the benchmark values $\tilde{\alpha}=\pm$1 km$^2$, $\gamma=\pm$1.  
Note that only $\rho_{\rm tot}$ and $\rho_{\rm rad}$ represent physical energy densities, while $\rho_{\phi}$ and $\rho_{\rm GB}$ are shown for illustrative purposes (in particular, $\rho_{\rm GB}$ can be negative and is plotted in absolute value).

\begin{figure*}[ht!]
\centering
\includegraphics[width=7.49cm,height=5cm]{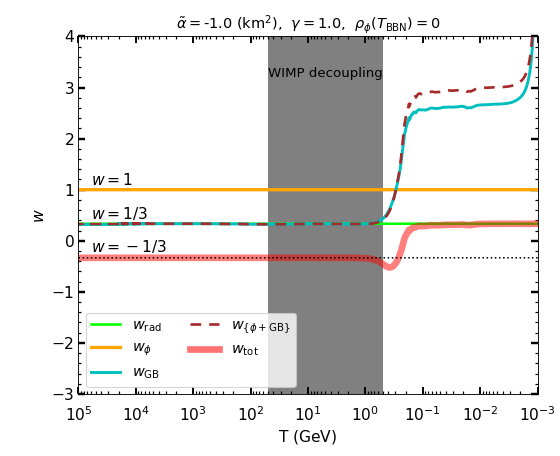}
\includegraphics[width=7.49cm,height=5cm]{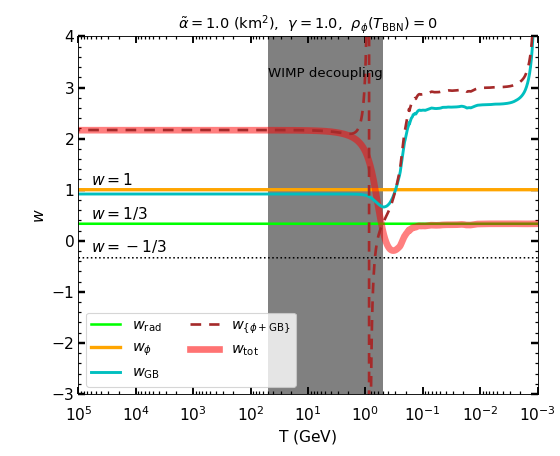}
\caption{Temperature evolution of the equation of state $w$ ($= p/\rho$) for the Universe and for different components.
\label{fig:EOS_phidot0}}
\end{figure*}

\begin{figure*}[ht!]
\centering
\includegraphics[width=7.49cm,height=5cm]{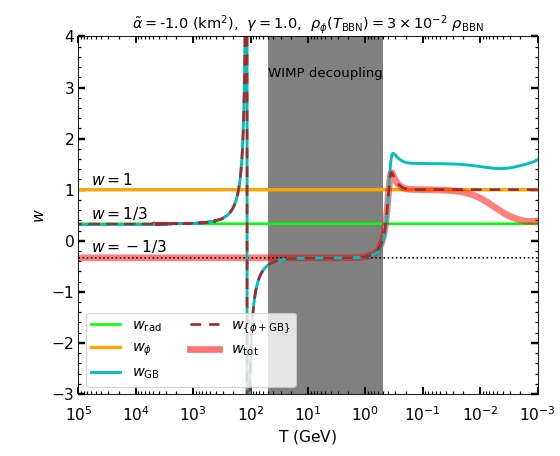}
\includegraphics[width=7.49cm,height=5cm]{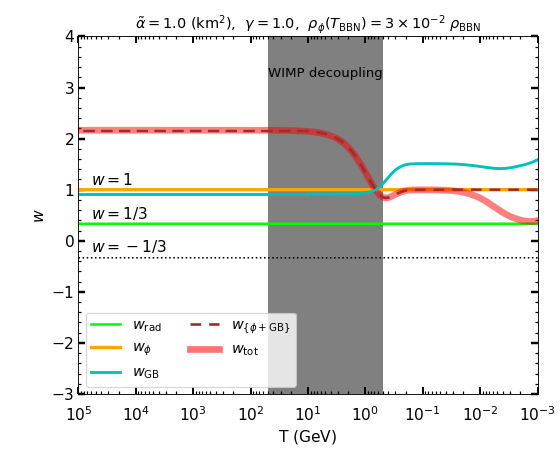}
\includegraphics[width=7.49cm,height=5cm]{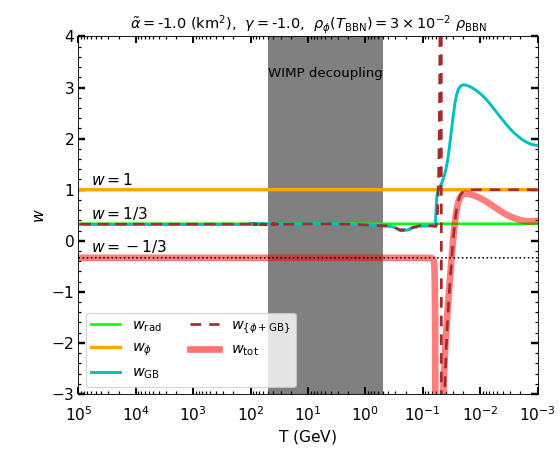}
\includegraphics[width=7.49cm,height=5cm]{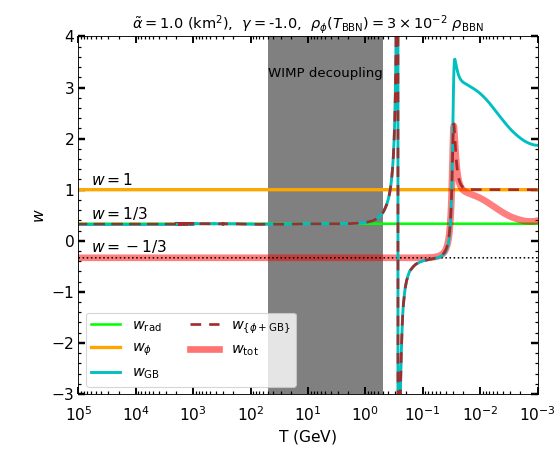}
\caption{The same as in Fig.~\ref{fig:EOS_phidot0} for $\rhoBBNT = 3 \times 10^{-2} \rho_{\rm BBN}$.}
\label{fig:EOS}
\end{figure*}

In Figs.~\ref{fig:EOS_phidot0} and \ref{fig:EOS} the equation of state of the Universe $w_{\rm tot}=p_{\rm tot}/\rho_{\rm tot}$ as well as those of the separate components $w_i=p_i/\rho_i$ are shown for the same benchmarks values of $\tilde{\alpha}$ and $\gamma$, with $\rhoBBNT=0$ and $\rhoBBNT =3\times 10^{-2}\rho_{\rm BBN}$,
respectively. Notice that in some cases $w_{\{\phi+GB\}}$ diverges because $\rho_{\{\phi+GB\}}$ changes sign.
However $\rho_{\{\phi+GB\}}$ maintains a smooth temperature behaviour. In particular the relation $\rho_i\propto T^{3(1+w_i)}$ only holds when $w_i$ is constant and for components that verify separately the continuity equation. In Figs.~\ref{fig:rho_phidot_0} and \ref{fig:EOS_phidot0}, where $\rho_{\phi}(T_{\rm BBN}$) = 0, plots for only $\gamma=1$ are provided since those for $\gamma=-1$ are identical. Moreover, two examples of the values reached by the enhancement parameter of Eq.~(\ref{eq:enhancement}) at the temperature $T= 50$ GeV are provided in Fig.~\ref{fig:GB_scan_A} in the $\tilde{\alpha}$–$\gamma$ plane, for $\rhoBBNT=0$ (left-hand plot) and $\rhoBBNT = 3\times 10^{-2}$ (right-hand plot).  
Notice that, as explained in the case of  Figs.~\ref{fig:rho_phidot_0} and~\ref{fig:EOS_phidot0}, the left hand plot in Fig.~\ref{fig:GB_scan_A} is symmetric under a change of sign of the $\gamma$ parameter.

The region shaded in gray between the two vertical solid lines in Figs.~\ref{fig:rho_phidot_0} to \ref{fig:EOS} represents the interval of temperatures where the WIMP decoupling takes place (500 MeV $<T<$ 50 GeV for the WIMP mass interval 10 GeV $<m_\chi<$ 1 TeV and $T_f\sim m_\chi/20$). In all the plots for this range of temperatures,  $\rho_{\rm tot}$ (the solid red line)  differs from $\rho_{\rm rad}$ (the dotted green line) that represents the evolution of the energy density in the case of radiation domination i.e., Standard Cosmology. 

\begin{figure*}[ht!]
\centering
\includegraphics[width=7.6cm,height=6.2cm]{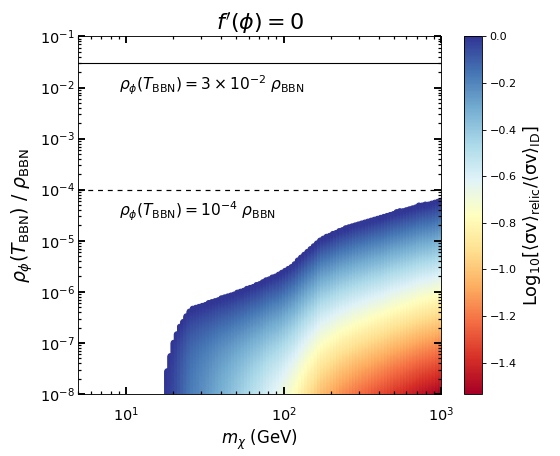}
\caption{Values of $\rhoBBNT$ favoured by WIMP 
indirect detection (ID) searches are shown by color coding 
for $m_{\chi}$ in the range 5 GeV - 1 TeV. 
Here both $f^{\prime}(\phi)$ and $V(\phi)$ are kept at zero. The three benchmark values of 
$\rhoBBNT$ used in this work are $\rhoBBNT = 0$, 
$10^{-4}$ $\rho_{\rm BBN}$ (shown by the black dashed line) 
and $3\times10^{-2}$ $\rho_{\rm BBN}$ (the black solid line). The color code of the shaded areas is the same as in Figs.~\ref{fig:GB_scan_relic_GW_1} and~\ref{fig:GB_scan_relic_GW_2} and explained in Section~\ref{sec:WIMP_constraints}.
\label{fig:mx_phidot_KE}}
\end{figure*}

\begin{figure*}[ht!]
\centering
\includegraphics[width=7.49cm,height=6.2cm]{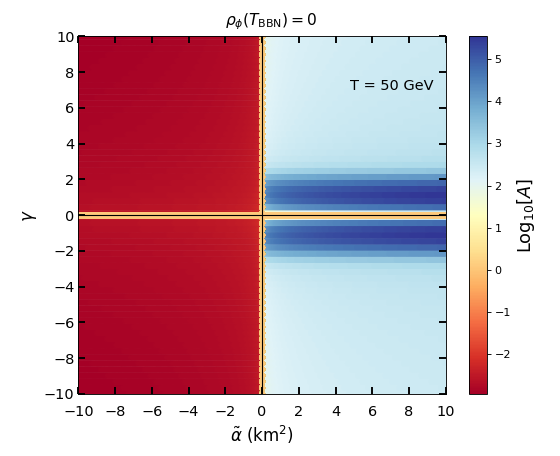}
\includegraphics[width=7.49cm,height=6.2cm]{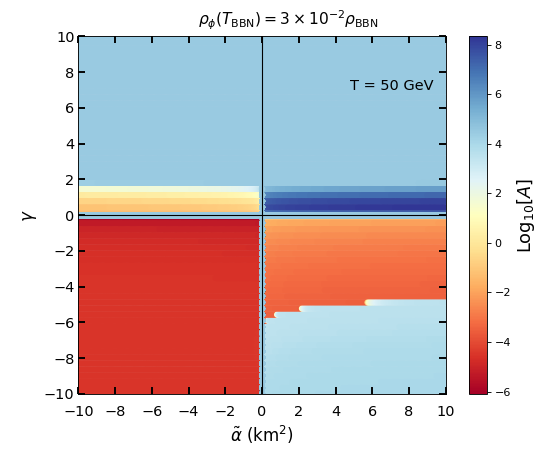}
\caption{Regions in the Gauss-Bonnet parameter space 
(spanned by $\tilde{\alpha}$ and $\gamma$) that correspond to different values of the enhancement factor $A$ ($= H / H_{\rm rad}$) at $\rm T = 50$ GeV are shown by different colors. The color bars indicate the values of ${\rm Log_{10}}[A]$. {\it Left column:} $\rho_{\phi} (T_{\rm BBN}) = 0$; 
{\it Right column:} $\rhoBBNT = 3 \times 10^{-2} \rho_{\rm BBN}$. Note that here we choose only the positive value of $\dot{\phi}_{\rm BBN}$ in $\rhoBBNT$. 
The corresponding results for the same but negative value of $\dot{\phi}$ can be obtained from these ones by flipping the sign of $\gamma$. 
}
\label{fig:GB_scan_A}
\end{figure*}

As a reference, let us consider the solution for $\tilde{\alpha}$ and/or $\gamma= 0$ (i.e., for a theory of the radiation and the scalar kinetic term with vanishing dEGB term). The radiation density evolves as $\rho_{\rm rad}\sim T^4$ over the whole temperature range. On the other hand, the generally complicated (photon) temperature dependence of the scalar field kinetic energy (kination) can be shown to be simplified as $\rho_{\phi}\sim T^6$ in the region where either the radiation or the scalar field (kination) dominates.  This is shown in purple in Fig.~\ref{fig:rho_phidot_neq0} (this case corresponds to radiation dominance in Fig.~\ref{fig:rho_phidot_0}).
As a consequence, for $\rhoBBNT \lesssim \epsilon\rho_{\rm BBN}$  the energy density of kination drives the Universe expansion for $T \gtrsim T_{\rm cross}=T_{\rm BBN}/\sqrt{\epsilon}$, with the enhancement factor $A(T)$ = $\sqrt{1+\epsilon (T/T_{\rm BBN})^2}$. For $\epsilon$ = $3\times 10^{-2}$ one gets $T_{\rm cross}\simeq$ 5.8 MeV and 80 $\lesssim A(T)\lesssim$ 8000 in the temperature range of WIMP decoupling, 500 MeV $\le T\le$ 50 GeV.
\noindent To predict the correct relic abundance such high enhancement factors require values of $\langle \sigma v\rangle_{\rm gal}$ = $\langle \sigma v\rangle_{\rm f}$ that exceed the bound discussed in Section~\ref{sec:indirect} unless $\rhoBBNT$ is much smaller than its upper bounds from BBN. This is shown in Fig.~\ref{fig:mx_phidot_KE}, where, for a vanishing dEGB term (kination only), the upper bound on $\rhoBBNT$ from WIMP indirect detection is plotted as a function of $m_\chi$ and compared to the two representative non-vanishing values that will be used in the quantitative analysis of Section~\ref{sec:WIMP_constraints}, $\rhoBBNT$ = $3\times 10^{-2}\rho_{\rm BBN}$ and $\rhoBBNT$ = $10^{-4}\rho_{\rm BBN}$. Fig.~\ref{fig:mx_phidot_KE} shows that in absence of the dEGB term both values are excluded. On the other hand, in Section~\ref{sec:WIMP_constraints} we will show that many combinations of $\tilde{\alpha}\ne0$, $\gamma\ne0$ are allowed for the same two values of $\rhoBBNT$. This clearly indicates that for such configurations the dEGB term plays a mitigating role on the kination dynamics, slowing down the speed of the scalar field evolution and reducing in this way the predicted values of the enhancement factor $A(T)$. Notice also that for $\tilde{\alpha}$ and/or $\gamma=0$ (kination only) $\rho_{\phi}$ vanishes at all temperatures if $\rhoBBNT=0$, while this is no longer true in presence of the dEGB term.  

Note that $T_{\rm cross}$ becomes higher as $\epsilon$ becomes smaller. Actually, when $\epsilon=0$, i.e.,  kination at $T_{\rm BBN}$ is zero, $\dot{\phi}(T)$ becomes identically zero, reducing to Standard Cosmology where the evolution of the Universe is simply given by radiation. Hence, the enhancement factor $A(T)\equiv 1$. This can be seen on the axes of the left hand plot of Fig.~\ref{fig:GB_scan_A}, where $\tilde{\alpha}\gamma=0$.

We now investigate the numerical solutions plotted in Figs.~\ref{fig:rho_phidot_0} and \ref{fig:rho_phidot_neq0}. Near $T=T_{\rm BBN}$ the dEGB term is negligible and radiation dominates. The Hubble parameter grows as $H\sim T^2$, while $\dot{\phi}\sim T^3$ and $\rho_{\phi}\sim T^6$, which grows faster than $\rho_{\rm rad}\sim T^4 $. So radiation domination is followed by an era dominated by kination at ~$T\gtrsim T_{\rm cross}\sim$ 5.8 MeV. This qualitative analysis shows consistency with Figs.~\ref{fig:rho_phidot_0} and \ref{fig:rho_phidot_neq0}. However, at higher temperatures the presence of the dEGB term has peculiar non-linear effects on the evolution of $\phi$ and $\dot{\phi}$ that significantly modify the energy density of the Universe and of the enhancement factor $A(T)$ compared to the scenario of simple kination. In particular $\rho_{\rm GB}$ quickly becomes of the same order of $\rho_{\phi}$, since at this stage its equation of state is much larger than any other component (see  Fig.~\ref{fig:EOS} and Appendix~\ref{app:w_GB}).
In order to understand what happens at higher temperatures we analyse Eq.~(\ref{Eq_frw3}):
in particular the largest values for the enhancement parameter are reached when $\dot{\phi}$ grows faster with $T$ than kination in the interval of temperatures relevant for the WIMP decoupling. 

The case $\dot{\phi}_{\rm BBN}=0$ is shown in Fig.~\ref{fig:rho_phidot_0}. The invariance by the transformation $\dot{\phi}_{\rm BBN}\rightarrow -\dot{\phi}_{\rm BBN}$,  $\gamma\rightarrow -\gamma$ implies that the results do not depend on the sign on $\gamma$ (except that $\dot{\phi}\rightarrow -\dot{\phi}$), so only $\gamma=1$ is shown, for both signs of $\tilde{\alpha}$. For $\tilde{\alpha}>0$ (right-hand plot) at $T_{\rm BBN}$ one has $V^{\prime}_{\rm GB}>0$ in the scalar field equation of motion, so that $\ddot{\phi}<0$ and $d\dot{\phi}/dT>0$. This means that $\dot{\phi}$ grows positive with $T$ (so that $3 H \dot{\phi}\ge 0$ and $\rho_{\rm GB}<0$) which implies that $d\phi/dT<0$ and that $\phi$ grows negative, with kination domination close to $T_{\rm BBN}$ with $3 H \dot{\phi}\gg V^{\prime}_{\rm GB}$. As $\phi$ grows negative with $T$ both $\rho_{\rm GB}$ and $V^{\prime}_{\rm GB}$ get suppressed by the exponential term $e^{\gamma\phi}$, so that $V^{\prime}_{\rm GB}$ does not affect the field evolution and $\rho_{\rm GB}$ is never competing with $\rho_{\phi}$ to drive the Universe expansion. As a consequence, at high temperature the energy of the Universe is driven by kination and as shown in the left-hand plot of Fig.~\ref{fig:GB_scan_A} the enhancement factor at high temperature is higher than unity. 
On the other hand, for $\tilde{\alpha}<0$ the left-hand plot
of Fig.~\ref{fig:rho_phidot_0} shows that $\rho_{\phi}$ stops growing and indeed in the left-hand plot of Fig.~\ref{fig:GB_scan_A} the enhancement factor turns out to be below unity.
This may appear surprising, since at $T_{\rm BBN}$ both $3 H \dot{\phi}$ and $V^{\prime}_{\rm GB}$ are negative, with $V^{\prime}_{\rm GB}$ enhanced by the $e^{\gamma\phi}$ term, so nothing apparently prevents a large growth of $\dot{\phi}$ with kination domination and large $A$ values. In particular, one now has $V^{\prime}_{\rm GB}<0$,  so $\ddot{\phi}>0$ and $d\dot{\phi}/dT<0$ which implies that $\dot{\phi}$ develops a negative value and also $3 H \dot{\phi}<0$, with $\rho_{\rm GB}<0$, $d\phi/dT>0$ and $\phi>0$.  So again, as in the previous case, $3 H \dot{\phi}$ and $V^{\prime}_{\rm GB}$ have the same sign and initially kination dominates. However, at variance with the case with $\tilde{\alpha}>0$, now $\rho_{\rm GB}$ is negative and exponentially enhanced, grows much faster than $\rho_{\phi}$ and reaches it very quickly. When this happens $\rho_{\{\phi+{\rm GB}\}}$ must remain positive, so a large cancellation between $\rho_{\rm GB}$ and $\rho_{\phi}$ is achieved, which is sufficient to flatten the temperature dependence of $\rho_{\rm tot} \simeq \rho_{\{\phi+{\rm GB}\}}$ and trigger an era of accelerated expansion (indeed, in Fig.~\ref{fig:EOS_phidot0} the equation of state of the Universe drops below -1/3). This changes the sign of the deceleration parameter $q$ and so of $V^{\prime}_{\rm GB}$ that depends on it (see Eq.~(\ref{Eq_frw3})), so that $\dot{\phi}$ starts decreasing, kination dominance stops and $A<1$. This back-reaction mechanism is discussed in more detail in Appendix~\ref{app:back_reaction} with the help of a semi-analytical approximation.  
Such results, obtained for the specific case $\tilde{\alpha} =\pm$1 km$^2$ and $\gamma = \pm 1$ are confirmed for a wider range of the parameters.
In particular the highest sensitivity is on $\gamma$, that enters in the exponent of $f(\phi)$. For $\tilde{\alpha}>0$  $f(\phi)$ is exponentially suppressed, so increasing $\gamma$ leads to a further suppression with no effect on the evolution of $\phi$.  On the other hand, for $\tilde{\alpha}<0$ the function $f(\phi)$ is exponentially enhanced, so increasing $\gamma$ accelerates the back-reaction effect, that shifts to lower temperatures, again with little effect on the evolution of $\phi$ at high temperatures. This is confirmed by the left-hand plot of Fig.~\ref{fig:GB_scan_A}, that provides for $\dot{\phi}_{\rm BBN}= 0$ a scan of the enhancement factor at $T$ = 50 GeV in the $\tilde{\alpha}$–$\gamma$ plane (note that in this plot the region close to the axes, $\tilde{\alpha}\gamma\rightarrow 0$, corresponds to Standard Cosmology, i.e. $A=1$). Indeed, from this plot one can qualitatively conclude that, at least at large-enough temperatures, $A(T)> 1$ for $\tilde{\alpha}>0$ and $A(T)<1$ for $\tilde{\alpha}<0$.

The situation for $\rhoBBNT \ne0$ is shown in Fig.~\ref{fig:rho_phidot_neq0}, where the value $\rhoBBNT$ = $3\times 10^{-2} \rho_{\rm BBN}$ is shown. Since $\dot{\phi}_{\rm BBN}>0$ in all the four possible cases $\phi$ grows negative with $T$ for $T\gtrsim T_{\rm BBN}$, the exponential $e^{\gamma\phi}$ is suppressed for $\gamma>0$ (upper plots) and enhanced for $\gamma<0$ (lower plots). Near $T\gtrsim T_{\rm BBN}$ the Gauss-Bonnet effect is small and $3 H \dot{\phi}>0$ and so $\ddot{\phi}_{\rm BBN}<0$ in all the four possible cases.
For $\tilde{\alpha}$, $\gamma>0$ (upper-right) and $V^{\prime}_{\rm GB}>0$ (same sign of $3 H \dot{\phi}$), $\ddot{\phi}$ never changes sign and kination dominates at high temperature since $\rho_{\phi}$ never stops growing. This case corresponds to the situation with the highest enhancement factor.
This is reflected in the right-hand plot of Fig.~\ref{fig:GB_scan_A}, where the value of the enhancement factor at $T$ = 50 GeV ranges between $10^4$ and $10^8$ for both $\tilde{\alpha}$ and $\gamma$ positive. On the other hand, for all the other three sign combinations of $\tilde{\alpha}$ and $\gamma$  kination stops dominating the Universe expansion at some stage, implying at the corresponding temperatures enhancement factors either moderate or less than unity. The two cases when $\tilde{\alpha}\gamma<0$ are easier to understand: now $V^{\prime}_{\rm GB}<0$, so eventually $\ddot{\phi}$ changes sign and $\dot{\phi}$ starts decreasing. This suppresses $\rho_{\phi}$ (that is quadratic in $\dot{\phi}$) more than $\rho_{\rm GB}$ (which is linear) so that kination dominance stops and a short era driven by $\rho_{\rm GB}$ (that at this stage is positive) starts, for which $A(T)>1$. However, this condition does not last much since at higher temperature $\dot{\phi}$ keeps decreasing and eventually crosses zero, so that both $\rho_{\phi}$ and $\rho_{\rm GB}$ vanish, leading the way to a short period of radiation domination, eventually followed by one driven by $\rho_{\rm rad}$ + $\rho_{\rm GB}<\rho_{\rm rad}$ (with $A<1$), when $\dot{\phi}$ and $\rho_{\rm GB}$ get negative and large enough in absolute value. 

\noindent At even larger $T$ the sign flip in $\dot{\phi}$ implies that the scalar field $\phi$ eventually becomes positive. What happens at this stage depends on the sign of $\gamma$. In particular, for $\gamma>0$ the dominance of $\rho_{\rm rad}$ + $\rho_{\rm GB}$ continues, because the exponential term in $\rho_{\rm GB}$ and  $V^{\prime}_{\rm GB}$ remains large. However, for $\gamma<0$ the  $V^{\prime}_{\rm GB}$ term is exponentially suppressed and the evolution of $\dot{\phi}$ is again only driven by $3 H \dot{\phi}$, triggering another period dominated by kination with $A>1$. 

\noindent Indeed, in the two panels with $\tilde{\alpha}\gamma<0$  the enhancement factor $A(T)$ is in general small or moderate, corresponding to the phase when the expansion of the Universe is driven by $\rho_{\rm rad}$ + $\rho_{\rm GB}<\rho_{\rm rad}$. However, for $\gamma<0$ larger values of $|\gamma|$ drive $A$ above unity. This happens because for $\gamma>0$  a larger $\gamma$ slows down the scalar field evolution delaying the epoch dominated by $\rho_{\rm GB}$ to a higher range of temperatures that includes $T$ = 50 GeV. On the other hand, when $\gamma<0$ and large enough in absolute value the scalar field evolution is faster, anticipating to $T$ = 50 GeV the second epoch of kination dominance triggered at high temperatures
when $\phi$ becomes positive.

The final case that remains to be discussed is that with both $\tilde{\alpha}$ and $\gamma$ negative, which corresponds to the bottom-left plot of Fig.~\ref{fig:rho_phidot_neq0}.  As confirmed by the right-hand plot of Fig.~\ref{fig:GB_scan_A}, in this case the evolution of the enhancement factor turns out to be below unity. 
This case is similar to the left-hand plot of Fig.~\ref{fig:rho_phidot_0} with $\rho_{\phi}(T_{\rm BBN})=0$ and $\tilde{\alpha}<0$. In fact also in this case at $T_{\rm BBN}$ the two quantities $V^{\prime}_{\rm GB}$ and $3 H \dot{\phi}$ have the same sign, with $\rho_{\rm GB} < 0$ and both GB terms exponentially 
enhanced. As a consequence, just above $T_{\rm BBN}$ an epoch of kination domination starts, but very quickly $\rho_{\rm GB} < 0$ reaches the level of $\rho_{\phi}$ and the Universe expansion is driven by $\rho_{\rm tot}$ = $\rho_{\{\phi + {\rm GB}\}}$, with a level of cancellation between $\rho_{\rm GB}$ and $\rho_{\phi}$ sufficient to flatten the temperature dependence of $\rho_{\rm tot}$ and trigger an era of accelerated expansion (indeed, in Fig.~\ref{fig:EOS} the equation of state of the Universe drops below -1/3).

\subsection{Constraints on the dEGB scenario}
\label{sec:late_constraints}

In this Section we will discuss the bounds on the dEGB scenario that can be obtained from the GW signals produced in compact binary mergers and those from WIMP indirect detection that are the main topic of our analysis.

Other classes of bounds that are usually applied on extensions of GR but that do not impose competitive constraints on the dEGB scenario include tests of gravity within the Solar System~\cite{Sotiriou:2006pq},  deviations of Kepler's formula for the motion of binary-pulsar systems~\cite{Yagi:2015oca} and constraints on dipole radiation emission from binary pulsars~\cite{Yagi:2015oca}.

\subsubsection{Black hole and neutron star binaries}
\label{sec:BH_NS}

The existence of black hole solutions in the 4D effective superstring action in presence of Gauss-Bonnet quadratic curvature terms~\cite{Kanti:1995vq} has triggered the study of the constraints that can be obtained on dEGB gravity from the observation of Gravitational Waves from BH-BH and BH-NS merger events. In particular, the waveforms obtained by modelling the different phases of compact binaries (inspiral \cite{Yagi:2011xp}, merger \cite{Witek:2018dmd} and ringdown \cite{Blazquez-Salcedo:2016enn}) in the presence of dEGB gravity can be compared to the data and constraints consistent with the statistical noise on the size of the dEGB deviation from standard GR can be obtained.

Following this procedure, it is already possible to use the data from the LIGO-Virgo Collaboration~\cite{LIGOScientific:2018mvr} to put constraints on deviations from GR. For a list of the events from which this has been carried out please check \cite{Nair:2019iur, Okounkova:2020rqw, Wang:2021jfc, Perkins:2021mhb, BH-NS_GB_2022}. From these studies it is possible to obtain a constraint on the GB term, $\alpha_{\text{GB}}$ in the notation of \cite{BH-NS_GB_2022}, of the order of $\alpha^{1/2}_{\text{GB}}\leq \mathcal O (2 \ \rm{km})$. In particular, for our analysis we use the value $\alpha^{1/2}_{\text{GB}} \leq  1.18\  \rm{km}$ \cite{BH-NS_GB_2022}.

Due to the Universe expansion even if $\dot{\phi}(T_{\rm BBN})\ne$ 0  the evolution of the scalar field $\phi$ eventually freezes at some asymptotic temperature $T_L\ll T_{\rm BBN}$ to a constant background value $\phi(T_L)$, implying no departure from GR at the cosmological level for $T<T_L$. On the other hand, in the vicinity of a BH or a NS the density profile of the scalar field is distorted compared to $\phi(T_L)$, leading to a local departure from GR that can modify the GW signal if the stellar object is involved in a merger event. Near the black hole or the neutron star the distortion of the scalar field is small and the dEGB function $f(\phi)$ can be expanded up to the linear term in the small perturbation $\Delta \phi$ around the asymptotic value $\phi(T_L)$ of the scalar field at large distance~\cite{BH-NS_GB_2022}:  
\begin{equation}
f(\phi) = f(\phi(T_L)) + f'\left(\phi(T_L)\right) \Delta \phi + {\cal O}( (\Delta \phi)^2).
\end{equation}
\noindent In this way the constraints from compact binary mergers is expressed in terms of $f'\left(\phi(T_L)\right)$:

\begin{equation}
|f'\left(\phi(T_L)\right)|
\leq \sqrt{8 \pi} \hspace{0.7mm} \alpha^{\rm max}_{\rm GB} , 
\label{eq:GW_constraints}
\end{equation}

\noindent with $\alpha^{\rm max}_{\rm GB} = (1.18)^2$ $\rm km^2$~\cite{BH-NS_GB_2022}. Note that the extra factor of $\sqrt{8 \pi}$ in the equation above
is due to the conversion from the units of Ref.~\cite{BH-NS_GB_2022} (that use $G = c = 1$)  to those of the present work (where $\kappa = 8 \pi G = c = 1$). 

At $T\lesssim T_{\rm BBN}$ $V^{\prime}_{\rm GB}$ is completely negligible and the scalar field equation is homogeneous, $\ddot{\phi}+3 H\dot{\phi}$ = $a^3 d/dt(a^3 \dot{\phi})= 0$. This implies that if $\dot{\phi}(T_{\rm BBN})= 0$ one has  $\dot{\phi}= 0$ also below $T_{\rm BBN}$ and as a consequence Eq.(\ref{eq:GW_constraints}) can be directly used with $\phi(T_L)$ = $\phi(T_{\rm BBN})$ to put constraints on the $\tilde{\alpha}$ and $\gamma$ parameters. 
The regions of  the $\tilde{\alpha}$–$\gamma$ parameter space that are disallowed by the ensuing constraint correspond to the hatched areas in the left-hand plots of
Fig.~\ref{fig:GB_scan_relic_GW_1}. 

However, if $\dot{\phi}(T_{\rm BBN})\ne 0$ 
one needs to consider the residual evolution of $\phi$ below $T_{\rm BBN}$ to calculate the value of the field $\phi(T_L)$ to be used in Eq.~(\ref{eq:GW_constraints}). 
In order to do so the derivative of $\phi$ with respect to the temperature can be expressed as:
\begin{equation}
\frac{d\phi}{dT} = \dot{\phi} \frac{dt}{dT} \simeq - \frac{\dot{\phi}}{H T} ,
\label{eq:dphi_dT}
\end{equation}
where in the last equality Eq.~(\ref{eq:T-t}) with $\beta\sim1$ has been used. Then neglecting $V^{\prime}_{\rm GB}$ the evolutions of $\dot{\phi}$ and $H$ below $T_{\rm BBN}$ are simply: 
\begin{eqnarray}
\dot{\phi}(T) &\simeq& \dot{\phi}_{\rm BBN} \left( \frac{T}{T_{\rm BBN}} \right)^3,
\\
H(T) &\simeq& H_{\rm BBN} \left( \frac{T}{T_{\rm BBN}} \right)^2 ,
\end{eqnarray}
\noindent assuming that the Universe expansion is driven by radiation at $T_L$.
Plugging the two expressions above in Eq.~(\ref{eq:dphi_dT}) one gets: 
\begin{equation}
\frac{d\phi}{dT} = - \frac{\dot{\phi}_{\rm BBN}}{H_{\rm BBN} T_{\rm BBN}} ,
\end{equation}
which upon integration from $T_{\rm BBN}$ to  $T_L$ yields:
\begin{eqnarray}
\phi(T_L) - \phi(T_{\rm BBN}) &=& - \frac{\dot{\phi}_{\rm BBN}}{H_{\rm BBN} T_{\rm BBN}} 
(T_L - T_{\rm BBN})
\nonumber\\
&=& \frac{\dot{\phi}_{\rm BBN}}{H_{\rm BBN} T_{\rm BBN}} (T_{\rm BBN} - T_L) 
\nonumber\\
&\simeq& \frac{\dot{\phi}_{\rm BBN}}{H_{\rm BBN}} \hspace{10mm} 
[{\rm for} \hspace{1.5mm} T_L << T_{\rm BBN}] . 
\end{eqnarray} 
So the value of $\phi(T_L)$ that must be used in Eq.~(\ref{eq:GW_constraints}) is, finally:
\begin{equation}
\phi(T_L) \simeq \phi_{\rm BBN} + \frac{\dot{\phi}_{\rm BBN}}{H_{\rm BBN}}.
\label{eq:phi_late}
\end{equation}

\noindent Using this Eq.~(\ref{eq:GW_constraints}) becomes:
\begin{equation}
|\tilde{\alpha} \gamma e^{\gamma \frac{\dot{\phi}_{\rm BBN}}{H_{\rm BBN}}}| 
\leq \sqrt{8 \pi} \hspace{0.7mm} \alpha^{\rm max}_{\rm GB} , 
\label{eq:GW_constraints_2}
\end{equation}

\noindent with $\tilde{\alpha}$ defined in Eq.~(\ref{eq:alpha_tilde}).

The expression above is used to determine the hatched excluded regions in the right-hand plot of Fig.~\ref{fig:GB_scan_relic_GW_1} and in Fig.~\ref{fig:GB_scan_relic_GW_2}  for $\rho_{\phi} (T_{\rm BBN}) = 3 \times 10^{-2} \rho_{\rm BBN}$ and for $\rho_{\phi} (T_{\rm BBN}) = 10^{-4} \rho_{\rm BBN}$, respectively.

\subsubsection{WIMP indirect detection}
\label{sec:WIMP_constraints}

In this Section we identify the regions of the Gauss-Bonnet parameter space that are favoured or disfavoured by WIMP dark matter. The favoured parameter regions are those for which 
the predicted  WIMP relic density falls within the observational range, $\Omega_{\chi}h^2 \simeq 0.12$, while at the same time the WIMP annihilation cross section in the halo of our Galaxy is compatible with indirect signals. In particular, for a given choice of the parameters we find the value ${\langle \sigma v \rangle}_{\rm relic}$ of ${\langle \sigma v \rangle}_{f}$ which yields $\Omega_{\chi}h^2 \simeq 0.12$ and compare it with the upper bound on the present annihilation cross section in the Milky Way ${\langle \sigma v \rangle}_{\rm ID}$ from DM indirect detection searches (see Table~\ref{tab:mx_sv_ID} and the discussion in Section~\ref{sec:indirect}). In order to do so we consider an $s$-wave annihilation cross section, for which ${\langle \sigma v \rangle}_{\rm gal}={\langle \sigma v \rangle}_{f}$.

\begin{figure*}[h!]
\centering
\includegraphics[width=7.49cm,height=5.75cm]{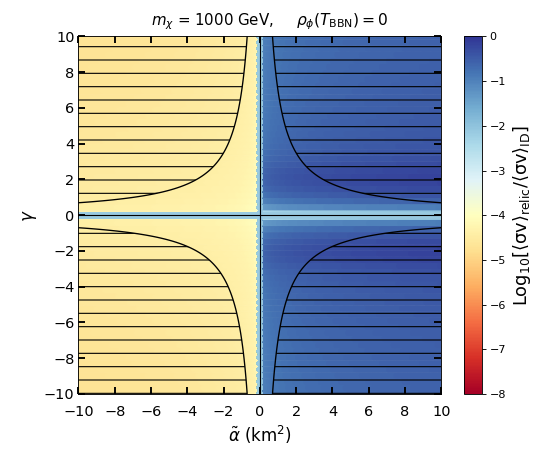}
\includegraphics[width=7.49cm,height=5.75cm]{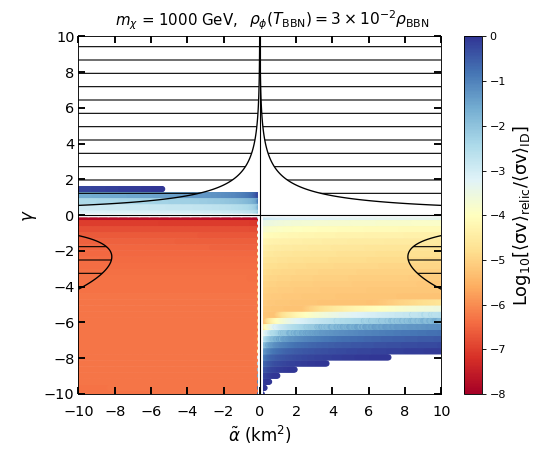}
\includegraphics[width=7.49cm,height=5.75cm]{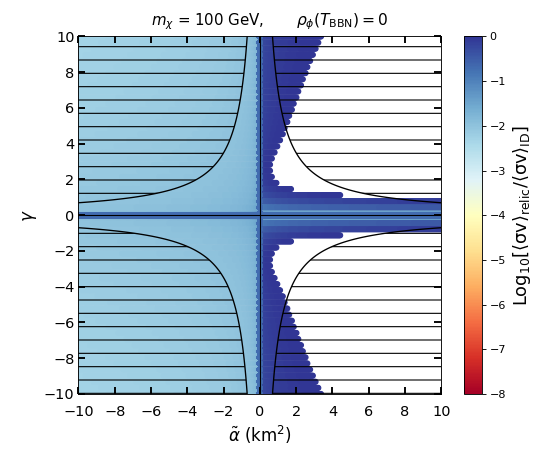}
\includegraphics[width=7.49cm,height=5.75cm]{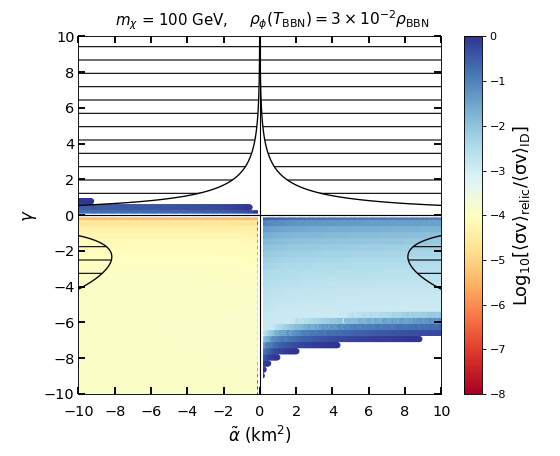}
\includegraphics[width=7.49cm,height=5.75cm]{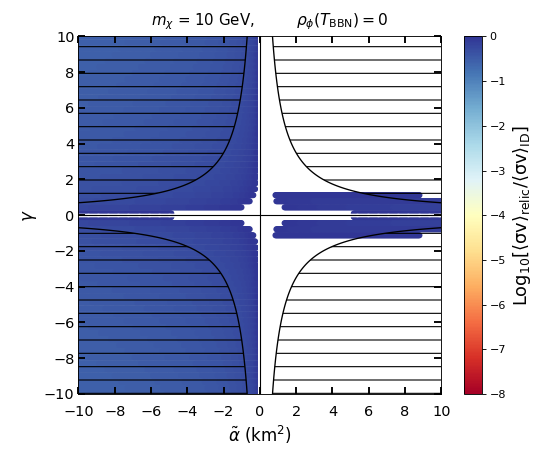}
\includegraphics[width=7.49cm,height=5.75cm]{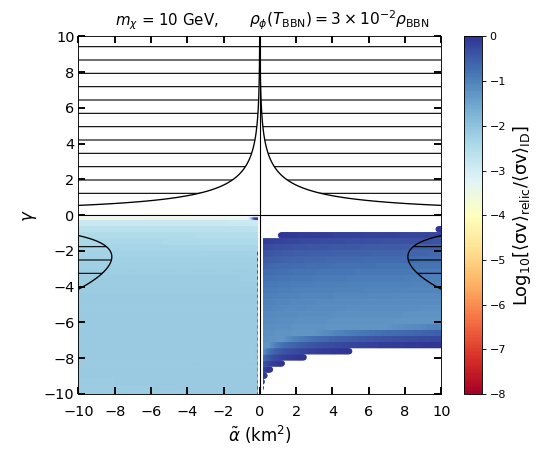}
\caption{
DEGB parameter space where the WIMP relic abundance corresponds to the observed one, i.e. $\Omega_{\chi}h^2 \simeq 0.12$. 
The color code refers to the ratio $\rm Log_{10} \left[ {\langle \sigma v \rangle}_{\rm relic} / {\langle \sigma v \rangle}_{\rm ID} \right]$.  
The white regions are excluded by WIMP indirect searches, while the hatched ones are ruled out by the detection of GW from compact binary mergers~\cite{BH-NS_GB_2022}. 
{\bf Top:} $m_{\chi} = 1$ TeV; {\bf middle:} $m_{\chi} = 100$ GeV; {\bf bottom:} $m_{\chi} = 10$ GeV. 
{\it Left column:} $\rhoBBNT = 0$; 
{\it Right column:} $\rhoBBNT = 3 \times 10^{-2} \rho_{\rm BBN}$. 
}
\label{fig:GB_scan_relic_GW_1}
\end{figure*}

\begin{figure*}[ht!]
\centering
\includegraphics[width=7.49cm,height=6.2cm]{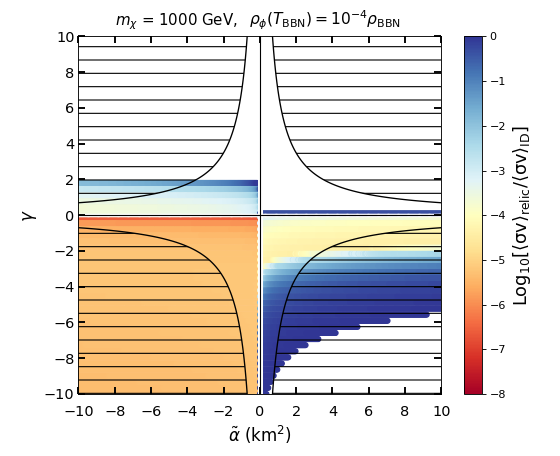}\\
\includegraphics[width=7.49cm,height=6.2cm]{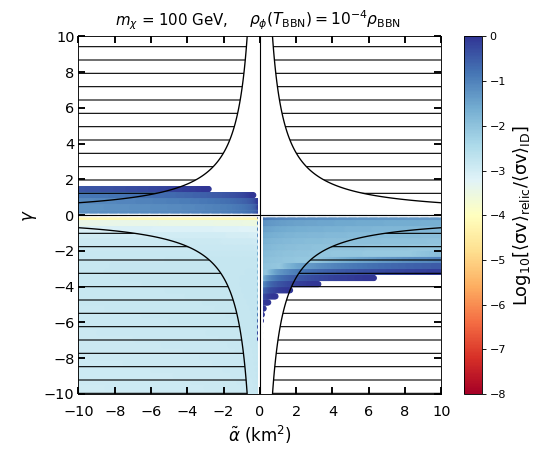}\\
\includegraphics[width=7.49cm,height=6.2cm]{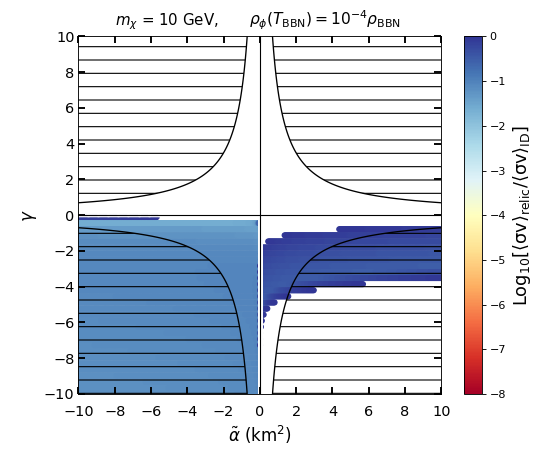}
\caption{The same of Fig. \ref{fig:GB_scan_relic_GW_1} for
$\rho_{\phi} (T_{\rm BBN}) = 10^{-4} \rho_{\rm BBN}$.}
\label{fig:GB_scan_relic_GW_2}
\end{figure*}

In Figs.~\ref{fig:GB_scan_relic_GW_1} and \ref{fig:GB_scan_relic_GW_2} the different colors in the $\tilde{\alpha}$–$\gamma$ plane correspond to values of logarithmic ratio, $\rm Log_{10} \left[ {\langle \sigma v \rangle}_{\rm relic} / {\langle \sigma v \rangle}_{\rm ID} \right]$. In particular, Fig.~\ref{fig:GB_scan_relic_GW_1} shows the two cases 
$\rhoBBNT = 0$ (left column) and $3 \times 10^{-2} \rho_{\rm BBN}$ (right column), while Fig.~\ref{fig:GB_scan_relic_GW_2} corresponds to $\rhoBBNT = 10^{-4} \rho_{\rm BBN}$. In both figures we present our results for three values of the WIMP mass, $m_{\chi} =$ 1000 GeV (top panels), 100 GeV (middle panels) and 10 GeV (bottom panels). The white regions of the parameter space have ${\langle \sigma v \rangle}_{\rm relic} / {\langle \sigma v \rangle}_{\rm ID} > 1$ and are disfavoured by indirect searches. In particular, the regions of parameters in dark blue are within one order of magnitude of the present sensitivity, $1/10 \lesssim {\langle \sigma v \rangle}_{\rm relic} / {\langle \sigma v \rangle}_{\rm ID} \lesssim 1$ and could be within the reach of future observations. As already pointed out, the results for negative values of $\dot{\phi}_{\rm BBN}$ can be obtained by flipping the sign of $\gamma$. The plots in the left column of Fig.~\ref{fig:GB_scan_relic_GW_1} show that the standard cosmological scenario is modified also for $\rhoBBNT=0$, i.e. irrespective on the boundary conditions of the scalar field, as long as the $\tilde{\alpha}$ and $\gamma$ parameters are non-vanishing. As confirmed by their top-down symmetry the results for $\rhoBBNT=0$ do not depend on the sign of $\gamma$. 

As already anticipated in the discussion of Section~\ref{sec:numerical} the highest values of the enhancement factor $A$ are found for $\tilde{\alpha}$ and $\gamma$ positive and this is confirmed by the plots of Figs.~\ref{fig:GB_scan_relic_GW_1} and \ref{fig:GB_scan_relic_GW_2}, where the corresponding region is completely excluded for all three values of $m_{\chi}$, unless $\dot{\phi}_{\rm BBN}=0$. As far as the latter case is concerned, a comparison between the plot for $m_{\chi}=$1 TeV and $\dot{\phi}_{\rm BBN}=0$ in Fig.~\ref{fig:GB_scan_relic_GW_1} and that of the corresponding enhancement factor $A$ at $T=$ 50 GeV (i.e. $m_\chi/20$) at the left-hand plot of Fig.~\ref{fig:GB_scan_A} shows that, when $\tilde{\alpha}$ and $\gamma$ are both positive, values of the enhancement factor as high as 10$^5$ do not drive ${\langle \sigma v \rangle}_{\rm gal}$ beyond its upper bound. This shows that the value of ${\langle \sigma v \rangle}_{\rm gal}$ does not scale directly with $A$ and is a clear indication of the mitigating effect that the post-freeze-out WIMP annihilation process discussed in Section~\ref{sec:relic} can have on the relic density and that is expected when the temperature evolution of $H$ is faster than in the standard case.           

Another feature that is worth noticing at this stage is that
the cases for $m_\chi=$ 10 GeV and $\dot{\phi}_{\rm BBN}\ne0$ in Figs.~\ref{fig:GB_scan_relic_GW_1} and \ref{fig:GB_scan_relic_GW_2} differ from those at higher values of $m_\chi$, in that no allowed regions are found for $\tilde{\alpha}<0$ and $\gamma>0$. This is due to the fact that the plots of Figs.~\ref{fig:GB_scan_relic_GW_1} and \ref{fig:GB_scan_relic_GW_2} depend on $m_\chi$ both through the shift of the decoupling temperature $T_f\sim m_\chi/20$ and because the experimental bounds on ${\langle \sigma v \rangle}_{\rm gal}$ depend on the WIMP mass. In particular, as can be seen from Table~\ref{tab:mx_sv_ID}, at low WIMP masses the present bounds on ${\langle \sigma v \rangle}_{\rm gal}$ have already reached the standard value 3$\times$10$^{-26}$ cm$^3s^{-1}$, i.e. they already exclude the standard scenario. In this case, at variance with what happens for higher WIMP masses, a modified cosmological scenario such as the dEGB model discussed in the present paper is actually {\it required} to reconcile the indirect detection bounds with the observed relic density. 

We conclude by noticing that in Figs.~\ref{fig:GB_scan_relic_GW_1} and \ref{fig:GB_scan_relic_GW_2} the hatched regions excluded by the late-time constraints from compact binary mergers discussed in Section~\ref{sec:BH_NS} are nicely complementary to those from WIMP indirect detection, with cases allowed/excluded by both bounds or excluded by only one of the two.

\section{Conclusions}
\label{sec:conclusions}
In the present paper we have applied the physics of WIMP decoupling to probe Cosmologies in a dilatonic Einstein Gauss-Bonnet (dEGB) scenario where the Gauss–Bonnet term is non–minimally coupled to a scalar field with vanishing potential. In particular, in such scenario
standard cosmology is modified irrespective of the boundary conditions of the scalar field, as long as the non-minimal coupling is non-vanishing. 

We have put constraints on the model parameters using the fact that in a modified cosmological scenario the WIMP annihilation cross section at freeze-out $\langle \sigma v\rangle_f$ required to predict the correct relic abundance is modified compared to the standard value 3$\times$10$^{-26}$ cm$^3s^{-1}$ and this can drive the WIMP annihilation cross section in the halo of our Galaxy $\langle \sigma v\rangle_{\rm gal}$ beyond the bounds from DM indirect detection searches, when $\langle \sigma v\rangle_f$ = $\langle \sigma v\rangle_{\rm gal}$. On the other hand, at low WIMP masses the present bounds on ${\langle \sigma v \rangle}_{\rm gal}$ have already reached the standard value and already exclude the standard scenario, so that a modified cosmological scenario such as the one discussed in the present paper is required to reconcile the WIMP indirect detection bounds with the observed relic density. In our analysis we assumed WIMPs that annihilate to SM particles through an s-wave process.

At fixed $\langle \sigma v\rangle_f$ the relic abundance prediction grows with the enhancement factor $A$, given by the ratio between the Hubble constant in the dEGB scenario and its standard value. In particular, for a vanishing dEGB term the value of $A$ is very large and incompatible with WIMP bounds unless $\dot{\phi}_{\rm BBN}$ is much lower than the present constraints. On the other hand, for the class of solutions that comply with WIMP indirect detection bounds we found that the dEGB term plays a mitigating role on the scalar field (kination) dynamics, slowing down the speed of its evolution and reducing $A$. For such slow $\dot{\phi}$ solutions we observe that the corresponding boundary conditions at high temperature correspond asymptotically to an equation of state $w=-1/3$ and a vanishing deceleration parameter $q$. This implies that in this class of solutions the effect of dEGB at high $T$ is to add an accelerating term that exactly cancels the deceleration predicted by GR. In this regime the density of the Universe is driven by $\rho_{\rm tot}\simeq \rho_{\rm rad}+\rho_{\rm GB}$, with a large cancellation between $\rho_{\rm rad}$ and $\rho_{\rm GB}<0$.

The bounds that we found from WIMP indirect detection are nicely complementary to late-time constraints from compact binary mergers. This suggests that it could be interesting to use other Early Cosmology processes to probe the dEGB scenario. In particular, although from the phenomenological point of view the evolution of the Universe at temperatures much larger than the WIMP thermal decoupling are irrelevant to our analysis, it would be interesting to study the implications of the dEGB scenario on Inflation or on the evolution of density perturbations.

\section*{Acknowledgements}
 
This research was supported by the National Research Foundation of Korea (NRF) funded by the Ministry of Education through the Center for Quantum Space Time (CQUeST) with grant number 2020R1A6A1A03047877, by the Ministry of Science and ICT with grant number 2021R1F1A1057119 (SS), NRF-2020R1F1A1075472 (BHL), NRF-2022R1I1A1A01067336 (WL), and NRF-2021R1A4A2001897 (AB).
BHL thanks the hospitality of APCTP, where part of this work was done.
LY thanks the YST Program of the APCTP.

\appendix

\section{Comments on the semi-analytical solutions of the Friedmann equations}
\label{app:semi}
The evolution of the Friedmann equations in the dEGB scenario is highly non-linear, so a numerical discussion is mandatory in order to obtain reliable predictions. Anyway, using semi-analytical expression it is possible to get a qualitative insight on the class of solutions that are obtained numerically. In the following we wish to briefly clarify three specific issues: (i) the steep evolution of $\rho_{\rm GB}$ and its equation of state at $T_{\rm BBN}$; (ii) the back-reaction mechanism that stops the growth of $\dot{\phi}$ when the deceleration parameter changes sign; (iii) the asymptotic equation of state at high $T$ when $\rho_{\phi}$ is subdominant.

\subsection{Equation of state of the dEGB term at temperatures close to Big Bang Nucleosynthesis}
\label{app:w_GB}
The evolution of the scalar field and the scale factor is governed by Eqs.~(\ref{Eq_frw1}) and (\ref{Eq_frw3}) which are reported here for convenience:
\begin{eqnarray}
&& H^2 = \frac{\kappa}{3} \left( {\rho}_{\{\phi + \rm GB\}} +\rho_{\rm rad}\right) \,,\\
&&{\ddot \phi} + 3H {\dot \phi} + V^{\prime}_{\rm GB}=0\,, \label{eqn_phi}
\end{eqnarray}
where $V^{\prime}_{\rm GB}\equiv -f' R^{2}_{\rm GB}= -24 f' H^2({\dot H} + H^2)=24 \tilde{\alpha}\gamma e^{\gamma\phi} q H^4 \, .$
Near $T=T_{\rm BBN}$ the Universe expansion is dominated by radiation $H\simeq\frac{1}{2t}\sim T^2$, while the GB term in the scalar field equation is negligible, so that $\dot{\phi}\sim t^{-3/2}\sim T^3$. With our choice of the initial condition $\dot{\phi}_{\rm BBN} \ge 0$ and ${\phi}=0$, the scalar field behavior is $\ddot{\phi}\le 0$ and $\dot{\phi}$ as well as ${\phi}$ grow fast at higher temperature. In particular, the scalar energy density $\rho_{\phi}=\frac{1}{2}\dot{\phi}^2$ grows $\rho_{\phi}\sim T^6$, faster than $\rho_{\rm rad}\sim T^4 $ for higher $T$.
Recall that the formal energy density for the Gauss-Bonnet 
$\rho_{\rm GB} = -24f^\prime{\dot{\phi}} H^3$ can be either positive or negative. In any case, neglecting in Eq.~(\ref{eq:rho_GB}) the $T$ dependence of $f^{\prime}(\phi(T))$, its absolute magnitude will grow as $\rho_{\rm GB} \sim T^9$, much faster than both $\rho_{\rm rad}\sim T^4 $ and $\rho_{\phi}\sim T^6$.  Figs.~\ref{fig:rho_phidot_0} and \ref{fig:rho_phidot_neq0} are consistent with this qualitative analysis. Moreover, using also (\ref{eq:p_GB}) one obtains:

\begin{equation}
w_{\rm GB}=\frac{p_{\rm GB}}{\rho_{\rm GB}}=-\frac{1}{3}\frac{d \ln(\dot{f}H^2)}{H dt}
-\frac{2}{3} =\frac{1}{3}\frac{d \ln(\dot{f}H^2)}{d\ln T}
-\frac{2}{3}.
\end{equation}

 \noindent and so $w_{\rm GB}(T_{\rm BBN})=5/3$ (in good agreement with Fig.~\ref{fig:EOS}). Notice that the evolution of $\rho_{\rm GB}$ is coupled to that of $\rho_{\phi}$, so $\rho_{\rm GB}\ne T^{3(1+w_{\rm GB})}=T^8$. 
\noindent 

\subsection{Back-reaction on the scalar field evolution from
the change of sign of the deceleration parameter} 
\label{app:back_reaction}
As explained in \ref{app:w_GB}, in absence of the Gauss-Bonnet term the kinetic energy of the scalar field grows with temperature as $\rho_{\phi}\sim T^6$ and eventually dominates at higher temperature. 
In the allowed parameter space discussed in Section \ref{sec:WIMP_constraints} the dEGB term is then instrumental in mitigating the growth of $\rho_{\phi}$. 

As explained in Section~\ref{sec:numerical} the suppression of the growth of $\dot{\phi}$ is relatively easy to understand when
close to $T_{\rm BBN}$ in Eq.(\ref{eqn_phi}) the two terms $V^{\prime}_{\rm GB}$ and $3 H \dot{\phi}$ have opposite sign, something that happens when $\rho_{\phi}(T_{\rm BBN})>0$ and
$\tilde{\alpha}\gamma<0$. On the other hand, naively one may expect no suppression of the scalar field kinetic energy either when $\rho_{\phi}(T_{\rm BBN})=0$ and $\tilde{\alpha}<0$ (left-hand plot if Fig.~\ref{fig:rho_phidot_0}) or when $\rho_{\phi}(T_{\rm BBN})>0$ and the product $\tilde{\alpha}\gamma$ is positive (bottom-left plot of Fig.~\ref{fig:rho_phidot_neq0}), since close to $T_{\rm BBN}$ in both cases $V^{\prime}_{\rm GB}$ and $3 H \dot{\phi}$ have the same sign. However, in Figs~\ref{fig:rho_phidot_0} and \ref{fig:rho_phidot_neq0} one observes numerically a suppression of the growth of $\dot{\phi}$ also in such circumstances.  

We wish here to provide a few comments on the non-linear back-reaction effect that is responsible of this behaviour.
As already pointed out $\rho_{\phi}$ quickly dominates over $\rho_{\rm rad}$ above $T_{\rm BBN}$. Soon after that also $|\rho_{\rm GB}|$ reaches the level of $\rho_{\phi}$, but since $\rho_{\rm GB}<0$ and the sum $\rho\sim\rho_{\phi}+\rho_{\rm GB}$ must remain positive a large cancellation arises between $\rho_{\phi}$ and $\rho_{\rm GB}$ (this is  observed both in Fig.~\ref{fig:rho_phidot_0} and~\ref{fig:rho_phidot_neq0} ). Since $|\rho_{\rm GB}|$ tracks $\rho_{\phi}$ very closely one can assume
$|\rho_{\rm GB}|\sim\rho_{\phi}$. Setting $\dot{\phi}\sim T^{\alpha}$ and parameterizing the additional $T$ dependence of $f^{\prime}$ as $f^{\prime}\sim T^{\xi}$ (with $\xi>0$ when $\gamma<0$), one can find a relation between the two parameters $\alpha$ and $\xi$ and the equation of state $w$ that drives the total density, $\rho\sim T^{3(1+w)}$:

\begin{equation}
\rho_{\phi}\sim T^{2\alpha} \sim |\rho_{\rm GB}| \sim T^{\alpha+\xi+\frac{9}{2}(1+w)}\rightarrow \alpha\sim \frac{9}{2}(1+w)+\xi.
\label{eq:back1}
\end{equation}

\noindent The same cancellation must happen in the pressure term, $p\sim p_{\phi}+p_{\rm GB}$:

\begin{equation}
p=\rho_{\phi}+p_{\rm GB}=\rho_{\phi}+\rho_{\rm GB}+8\frac{d(\dot{f}H^2)}{dt}-\frac{5}{3} \rho_{\rm GB}
\end{equation}
\noindent so that also the term $8d(\dot{f}H^2)/dt$ = $-8HT d(\dot{f}H^2)/dT$  must track closely 
$-5/3\rho_{\rm GB}=40 \dot{f} H^3$, or $Td(\dot{f}H^2)/dT \sim 5 \dot{f}H^2$. As a consequence:

\begin{equation}
\frac{d(\dot{f}H^2)}{\dot{f}H^2}\sim 5 \frac{dT}{T}\rightarrow f^{\prime}\dot{\phi}H^2 \sim T^{\xi+\alpha+3(1+w)}\sim T^5 \rightarrow \xi+\alpha+3(1+w)=5,
\label{eq:back2}
\end{equation}
\noindent and from the two relations ~(\ref{eq:back1}) and (\ref{eq:back2}) one gets:

\begin{equation}
\alpha=3-\frac{\xi}{5},\;\;\;w=-\frac{1}{3}-\frac{4}{15}\xi.
\end{equation}

The equation above shows that as $|\rho_{\rm GB}|$ approaches $\rho_{\phi}$ the equation of state $w$ turns smaller than -1/3, triggering a change of sign in the deceleration parameter $q=(1+3w)/2$.
This implies that also $V^{\prime}_{\rm GB}$ in the scalar field equation changes sign. Thanks to the exponential enhancement in $f$ eventually $V^{\prime}_{\rm GB}$ overcomes $3 H \dot{\phi}$, flipping the sign of $\ddot{\phi}$ and stopping the growth of $\dot{\phi}$.  

\subsection{Equation of state at high temperature for subdominant kination energy}
\label{app:w_low_phi_dot}

One can notice that in Figs.~\ref{fig:EOS_phidot0} and \ref{fig:EOS} whenever $\rho_{\phi}$ is subdominant at high temperature the corresponding equation of state of the Universe gets close to $w \sim $-1/3, i.e. $q=0$. 
This implies that in this class of solutions the effect of dEGB is to add an accelerating term that exactly cancels the deceleration predicted by GR (notice that in our model $V=0$).  

This is an asymptotic solution of the Friedmann equation that can be verified by solving for $\dot{H}$ in Eq.~(\ref{Eq_frw2}) and (\ref{Eq_frw3}):

\begin{equation}
    \dot{H}=-\frac{\kappa(
    8H^2f^{\prime\prime}\dot{\phi}^2-32H^3f^{\prime}\dot{\phi}+192H^6(f^{\prime})^2+p_{\rm rad}+\rho_{\rm rad}+\dot{\phi}^2
    )    
    }
    {2(
    96 \kappa (f^{\prime})^2 H^4 +8 \kappa f^{\prime}\dot{\phi}H +1
    )
    }.
\label{H_dot}
\end{equation}
In this specific scenario the density of the Universe is $\rho_{\rm tot}\simeq \rho_{\rm rad}+\rho_{\rm GB}\gg \rho_{\phi}$, with  $\rho_{\rm GB}$ closely tracking $\rho_{\rm rad}$.
We will write the terms in Eq.~(\ref{H_dot}) by expressing $f^{\prime}$, $f^{\prime\prime}=\gamma f^{\prime}$, ${\dot{\phi}}$ and $\kappa$ 
in terms of $\rho_{\rm GB}=-24f^{\prime} H^3$, $\rho_{\phi}=\frac{1}{2} {\dot{\phi}}^2$ and using $\kappa=\frac{3H^2}{\rho_{\rm tot}}$.  
Roughly speaking, $f^{\prime}$ is very large because it converts to $\rho_{\rm GB}$, while $\dot\phi$ is very small because it converts to $\rho_{\phi}$. Rewriting in terms of magnitude orders:
\begin{equation}
    \dot{H}=-\frac{\kappa\left(
    192(f^{\prime})^2H^6 +8f^{\prime\prime}\dot{\phi}^2H^2
    +\frac{4}{3}\left(\rho_{\rm rad}+\rho_{\rm GB}\right)+\dot{\phi}^2
                         \right)    
    }
    {2\left(
    96 \kappa (f^{\prime})^2 H^4 +8 \kappa f^{\prime}\dot{\phi}H +1
    \right)
    }.
\label{H_dot_order}
\end{equation}

\noindent 
In this specific scenario the density of the Universe is $\rho_{\rm tot}=\rho_{\rm rad}+\rho_{\rm GB}+\rho_{\phi}\simeq \rho_{\rm rad}+\rho_{\rm GB}\gg \rho_{\phi}$, with  $\rho_{\rm GB}$ closely tracking $\rho_{\rm rad}$.  In other words, the hierarchies are $\rho_{\rm rad} \, {\rm and}\, |\rho_{\rm GB}|\gg \rho_{\rm tot}\simeq \rho_{\rm rad}+\rho_{\rm GB}\gg \rho_{\phi}$. 
Among the three independent terms $\rho_{\rm rad}$, $\rho_{\phi}$, $\rho_{\rm GB}$ 
(and $\rho_{\rm tot}=\rho_{\rm rad} +\rho_{\rm GB} +\rho_{\phi} $) 
we can make two independent ratios, which we choose as 
$\left(\frac{\rho_{\phi}}{\rho_{\rm GB}}\right)$ and $\left(\frac{\rho_{\rm tot}}{\rho_{\rm GB}}\right)$. These two ratios are very small numbers, allowing us to expand quantities in power series of these. The ratio of these two small numbers $\left(\frac{\rho_{\phi}}{\rho_{\rm tot}}\right) \equiv r$ is also much smaller than 1 in our situation.
Finally the ratio of radiation energy to ${\rho_{\rm GB}}$ is written as $\left(\frac{\rho_{\rm rad}}{\rho_{\rm GB}}\right) = -1-\left(\frac{\rho_{\phi}}{\rho_{\rm GB}}\right)+\left(\frac{\rho_{\rm tot}}{\rho_{\rm GB}}\right)$. This shows the tracking of $\rho_{\rm GB}$ to $\rho_{\rm rad}$ with the difference in terms of those two small independent ratios.

\noindent Using the following expressions: 
\begin{eqnarray}
 2\times96 \kappa (f^{\prime})^2 H^4 &=&2\frac{96\kappa(f^{\prime}\dot{\phi}H^3)^2}{{\dot\phi}^2 H^2} = \frac{1}{2} \left(\frac{\rho_{\rm GB}}{\rho_{\rm tot}}\right) \left(\frac{\rho_{\rm GB}}{\rho_{\phi}}\right),
 \nonumber \\
2\times8 \kappa f^{\prime} \dot\phi H &=&\frac{1}{2}\left(\frac{\rho_{\rm GB}}{\rho_{\rm tot}}\right) \left(\frac{\rho_{\rm GB}}{\rho_{\phi}}\right) \left(-4\left(\frac{\rho_{\phi}}{\rho_{\rm GB}}\right)\right),
\end{eqnarray}
the denominator can be written as 
\begin{equation}
    {\rm Denominator}
    =\frac{1}{2} \left(\frac{\rho_{\rm GB}}{\rho_{\rm tot}}\right) \left(\frac{\rho_{\rm GB}}{\rho_{\phi}}\right)
    \left( 1-4\left(\frac{\rho_{\phi}}{\rho_{\rm GB}}\right) 
    +4\left(\frac{\rho_{\rm tot}}{\rho_{\rm GB}}\right) \left(\frac{\rho_{\phi}}{\rho_{\rm GB}}\right) \right).
    \label{Denominator}
\end{equation}

\noindent In the numerator the first two terms are:
\begin{eqnarray}
\kappa 192(f^{\prime})^2H^6  &=& \frac{H^2}{2} \left(\frac{\rho_{\rm GB}}{\rho_{\rm tot}}\right) \left(\frac{\rho_{\rm GB}}{\rho_{\phi}}\right), \nonumber \\
8\kappa f^{\prime\prime}{\dot\phi}^2H^2  &=& \frac{H^2}{2} \left(\frac{\rho_{\rm GB}}{\rho_{\rm tot}}\right)                 \left(\frac{\rho_{\rm GB}}{\rho_{\phi}}\right) 
  \left(  -2\gamma\sqrt{\frac{6r}{\kappa}} \left(\frac{\rho_{\phi}}{\rho_{\rm GB}}\right) 
  \right).  \label{num_larger}
\end{eqnarray}

\noindent The remaining terms are:
\begin{eqnarray}
\frac{4}{3}\kappa\left(\rho_{\rm rad}+\rho_{\rm GB}\right) &=& 4 H^2\left(\frac{\rho_{\rm GB}}{\rho_{\rm tot}}\right) 
\left(-1+\left(\frac{\rho_{\rm tot}}{\rho_{\rm GB}}\right) -\left(\frac{\rho_{\phi}}{\rho_{\rm GB}}\right) +1\right) \nonumber \\ 
&=& \frac{H^2}{2}\left(\frac{\rho_{\rm GB}}{\rho_{\rm tot}}\right) \left(\frac{\rho_{\rm GB}}{\rho_{\phi}}\right)
\left(8\left(\frac{\rho_{\phi}}{\rho_{\rm GB}}\right)\left(\frac{\rho_{\rm tot}}{\rho_{\rm GB}}\right)
    -8\left(\frac{\rho_{\phi}}{\rho_{\rm GB}}\right)^2 
\right), \nonumber \\
\kappa {\dot\phi}^2 &=& 
   \frac{H^2}{2}\left(\frac{\rho_{\rm GB}}{\rho_{\rm tot}}\right) \left(\frac{\rho_{\rm GB}}{\rho_{\phi}}\right)
   \left( 12\left(\frac{\rho_{\phi}}{\rho_{\rm GB}}\right)^2 \right). \label{num_smaller}
\end{eqnarray}

\noindent By summing Eqs.~(\ref{num_larger}) and (\ref{num_larger}), the numerator becomes:
\begin{equation}
    \frac{H^2}{2} \left(\frac{\rho_{\rm GB}}{\rho_{\rm tot}}\right) \left(\frac{\rho_{\rm GB}}{\rho_{\phi}}\right)  
  \left( 
    1 -2\gamma\sqrt{\frac{6r}{\kappa}} \left(\frac{\rho_{\phi}}{\rho_{\rm GB}}\right) 
    +8\left(\frac{\rho_{\phi}}{\rho_{\rm GB}}\right)\left(\frac{\rho_{\rm tot}}{\rho_{\rm GB}}\right) +4\left(\frac{\rho_{\phi}}{\rho_{\rm GB}}\right)^2
  \right).  \label{Numerator} 
\end{equation}

\noindent Finally, we get $\dot H$ in Eq.~(\ref{H_dot_order}) by combining Eqs.~(\ref{Denominator}) and (\ref{Numerator}):
\begin{eqnarray}
{\dot H}
&=&-{H^2}\frac{1 -2\gamma\sqrt{\frac{6r}{\kappa}} \left(\frac{\rho_{\phi}}{\rho_{\rm GB}}\right) 
+(8+4r)\left(\frac{\rho_{\phi}}{\rho_{\rm GB}}\right)\left(\frac{\rho_{\rm tot}}{\rho_{\rm GB}}\right) 
  }
{ 1-4 \left(\frac{\rho_{\phi}}{\rho_{\rm GB}}\right) 
+4\left(\frac{\rho_{\rm tot}}{\rho_{\rm GB}}\right) \left(\frac{\rho_{\phi}}{\rho_{\rm GB}}\right)}. 
\end{eqnarray} 

\noindent Keeping the leading term in the expansion:
\begin{equation}
\frac{\dot H}{H^2} =-(1+q) \simeq -\left( 1 
+ \left(4-2\gamma\sqrt{\frac{6r}{\kappa}}\right)\left(\frac{\rho_{\phi}}{\rho_{\rm GB}}\right)
                             \right), 
\end{equation} 
\noindent and finally, at leading order the deceleration is given by:
\begin{equation}
 q\simeq\left(4+2\gamma\sqrt{\frac{6r}{\kappa}}\right)\left(\frac{\rho_{\phi}}{\rho_{\rm GB}}\right)
 \simeq 4\left(\frac{\rho_{\phi}}{\rho_{\rm GB}}\right).   
\label{eq:q_asymptotic}
\end{equation}
In the last equation we dropped the $r$-dependent term since $r\ll 1$ and $\kappa=1$ in our setting. As a consequence, $q$ is a small non-vanishing negative number. Indeed, from the plots of Fig.~\ref{fig:rho_phidot_neq0} one can see that numerically the parameter $|\rho_{\phi}/\rho_{\rm GB}|$ can be extremely small in this regime. This implies that in this scenario the Universe at high temperature expands with a very tiny acceleration and equation of state $w\simeq$-1/3, eventually catching up Standard Cosmology and radiation dominance at much lower temperatures.
Moreover, from Eq.~(\ref{eq:q_asymptotic}) one has $q\times \rho_{\rm GB}/\rho_{\phi}\rightarrow$ 4, so that
\begin{equation}
    \frac{V^{\prime}_{\rm GB}}{3 H \dot{\phi}}=-\frac{1}{6}q \frac{\rho_{\rm GB}}{\rho_{\phi}}\rightarrow-\frac{2}{3} 
\end{equation}
    
\noindent and in this regime the scalar field equation becomes:
\begin{equation}
    \ddot{\phi}+H\dot{\phi}=0,
\end{equation}

\noindent  that implies $\rho_{\phi}\simeq T^2$, as confirmed explicitly in the plots of Fig.~\ref{fig:rho_phidot_neq0} where $\rho_{\phi}$ is subdominant.

It is worth noticing that, while, as shown in Fig.~\ref{fig:EOS}, this implies that both $\rho_{\rm rad}$ and $\rho_{\rm GB}$ have separate equation of state $w=1/3$, as explained above the equation of state for their combination $\rho_{\rm rad}+\rho_{\rm GB}$ is instead $-1/3$. This can be understood because the equation of state of the Universe $w=p/\rho$ is the result of a limit where both $p$ and $\rho$ have very large cancellations, i.e. $w\rightarrow 0/0$. 
\newpage

\begin{thebibliography}{10}
\bibitem{Clifton:2011jh}
T.~Clifton, P.~G. Ferreira, A.~Padilla and C.~Skordis, \emph{{Modified Gravity
  and Cosmology}},
  \href{https://doi.org/10.1016/j.physrep.2012.01.001}{\emph{Phys. Rept.}
  {\bfseries 513} (2012) 1--189},
  [\href{https://arxiv.org/abs/1106.2476}{{\ttfamily 1106.2476}}].

\bibitem{Nojiri:2010wj}
S.~Nojiri and S.~D. Odintsov, \emph{{Unified cosmic history in modified
  gravity: from F(R) theory to Lorentz non-invariant models}},
  \href{https://doi.org/10.1016/j.physrep.2011.04.001}{\emph{Phys. Rept.}
  {\bfseries 505} (2011) 59--144},
  [\href{https://arxiv.org/abs/1011.0544}{{\ttfamily 1011.0544}}].

\bibitem{Will:2014kxa}
C.~M. Will, \emph{{The Confrontation between General Relativity and
  Experiment}}, \href{https://doi.org/10.12942/lrr-2014-4}{\emph{Living Rev.
  Rel.} {\bfseries 17} (2014) 4},
  [\href{https://arxiv.org/abs/1403.7377}{{\ttfamily 1403.7377}}].

\bibitem{Horndeski:1974wa}
G.~W. Horndeski, \emph{{Second-order scalar-tensor field equations in a
  four-dimensional space}},
  \href{https://doi.org/10.1007/BF01807638}{\emph{Int. J. Theor. Phys.}
  {\bfseries 10} (1974) 363--384}.

\bibitem{Woodard:2015zca}
R.~P. Woodard, \emph{{Ostrogradsky's theorem on Hamiltonian instability}},
  \href{https://doi.org/10.4249/scholarpedia.32243}{\emph{Scholarpedia}
  {\bfseries 10} (2015) 32243},
  [\href{https://arxiv.org/abs/1506.02210}{{\ttfamily 1506.02210}}].

\bibitem{Tsujikawa:2013fta}
S.~Tsujikawa, \emph{{Quintessence: A Review}},
  \href{https://doi.org/10.1088/0264-9381/30/21/214003}{\emph{Class. Quant.
  Grav.} {\bfseries 30} (2013) 214003},
  [\href{https://arxiv.org/abs/1304.1961}{{\ttfamily 1304.1961}}].

\bibitem{Harko:2013gha}
T.~Harko, F.~S.~N. Lobo and M.~K. Mak, \emph{{Arbitrary scalar field and
  quintessence cosmological models}},
  \href{https://doi.org/10.1140/epjc/s10052-014-2784-8}{\emph{Eur. Phys. J. C}
  {\bfseries 74} (2014) 2784},
  [\href{https://arxiv.org/abs/1310.7167}{{\ttfamily 1310.7167}}].

\bibitem{Cicoli:2018kdo}
M.~Cicoli, S.~De~Alwis, A.~Maharana, F.~Muia and F.~Quevedo, \emph{{De Sitter
  vs Quintessence in String Theory}},
  \href{https://doi.org/10.1002/prop.201800079}{\emph{Fortsch. Phys.}
  {\bfseries 67} (2019) 1800079},
  [\href{https://arxiv.org/abs/1808.08967}{{\ttfamily 1808.08967}}].

\bibitem{Bahamonde:2017ize}
S.~Bahamonde, C.~G. B\"ohmer, S.~Carloni, E.~J. Copeland, W.~Fang and
  N.~Tamanini, \emph{{Dynamical systems applied to cosmology: dark energy and
  modified gravity}},
  \href{https://doi.org/10.1016/j.physrep.2018.09.001}{\emph{Phys. Rept.}
  {\bfseries 775-777} (2018) 1--122},
  [\href{https://arxiv.org/abs/1712.03107}{{\ttfamily 1712.03107}}].

\bibitem{Alexander:2019rsc}
S.~Alexander and E.~McDonough, \emph{{Axion-Dilaton Destabilization and the
  Hubble Tension}},
  \href{https://doi.org/10.1016/j.physletb.2019.134830}{\emph{Phys. Lett. B}
  {\bfseries 797} (2019) 134830},
  [\href{https://arxiv.org/abs/1904.08912}{{\ttfamily 1904.08912}}].

\bibitem{Banerjee:2020xcn}
A.~Banerjee, H.~Cai, L.~Heisenberg, E.~O. Colg\'ain, M.~M. Sheikh-Jabbari and
  T.~Yang, \emph{{Hubble sinks in the low-redshift swampland}},
  \href{https://doi.org/10.1103/PhysRevD.103.L081305}{\emph{Phys. Rev. D}
  {\bfseries 103} (2021) L081305},
  [\href{https://arxiv.org/abs/2006.00244}{{\ttfamily 2006.00244}}].

\bibitem{Odintsov:2018ggm}
S.~D. Odintsov and V.~K. Oikonomou, \emph{{The reconstruction of $f(\phi)R$ and
  mimetic gravity from viable slow-roll inflation}},
  \href{https://doi.org/10.1016/j.nuclphysb.2018.01.027}{\emph{Nucl. Phys. B}
  {\bfseries 929} (2018) 79--112},
  [\href{https://arxiv.org/abs/1801.10529}{{\ttfamily 1801.10529}}].

\bibitem{Nojiri:2006ri}
S.~Nojiri and S.~D. Odintsov, \emph{{Introduction to modified gravity and
  gravitational alternative for dark energy}},
  \href{https://doi.org/10.1142/S0219887807001928}{\emph{eConf} {\bfseries
  C0602061} (2006) 06}, [\href{https://arxiv.org/abs/hep-th/0601213}{{\ttfamily
  hep-th/0601213}}].

\bibitem{Hwang:1999gf}
J.-c. Hwang and H.~Noh, \emph{{Conserved cosmological structures in the one
  loop superstring effective action}},
  \href{https://doi.org/10.1103/PhysRevD.61.043511}{\emph{Phys. Rev. D}
  {\bfseries 61} (2000) 043511},
  [\href{https://arxiv.org/abs/astro-ph/9909480}{{\ttfamily
  astro-ph/9909480}}].

\bibitem{Satoh:2008ck}
M.~Satoh and J.~Soda, \emph{{Higher Curvature Corrections to Primordial
  Fluctuations in Slow-roll Inflation}},
  \href{https://doi.org/10.1088/1475-7516/2008/09/019}{\emph{JCAP} {\bfseries
  09} (2008) 019}, [\href{https://arxiv.org/abs/0806.4594}{{\ttfamily
  0806.4594}}].

\bibitem{Zwiebach:1985uq}
B.~Zwiebach, \emph{{Curvature Squared Terms and String Theories}},
  \href{https://doi.org/10.1016/0370-2693(85)91616-8}{\emph{Phys. Lett. B}
  {\bfseries 156} (1985) 315--317}.

\bibitem{Kanti:1995vq}
P.~Kanti, N.~E. Mavromatos, J.~Rizos, K.~Tamvakis and E.~Winstanley,
  \emph{{Dilatonic black holes in higher curvature string gravity}},
  \href{https://doi.org/10.1103/PhysRevD.54.5049}{\emph{Phys. Rev. D}
  {\bfseries 54} (1996) 5049--5058},
  [\href{https://arxiv.org/abs/hep-th/9511071}{{\ttfamily hep-th/9511071}}].

\bibitem{Cai:2001dz}
R.-G. Cai, \emph{{Gauss-Bonnet black holes in AdS spaces}},
  \href{https://doi.org/10.1103/PhysRevD.65.084014}{\emph{Phys. Rev. D}
  {\bfseries 65} (2002) 084014},
  [\href{https://arxiv.org/abs/hep-th/0109133}{{\ttfamily hep-th/0109133}}].

\bibitem{Guo:2010jr}
Z.-K. Guo and D.~J. Schwarz, \emph{{Slow-roll inflation with a Gauss-Bonnet
  correction}}, \href{https://doi.org/10.1103/PhysRevD.81.123520}{\emph{Phys.
  Rev. D} {\bfseries 81} (2010) 123520},
  [\href{https://arxiv.org/abs/1001.1897}{{\ttfamily 1001.1897}}].

\bibitem{Koh:2014bka}
S.~Koh, B.-H. Lee, W.~Lee and G.~Tumurtushaa, \emph{{Observational constraints
  on slow-roll inflation coupled to a Gauss-Bonnet term}},
  \href{https://doi.org/10.1103/PhysRevD.90.063527}{\emph{Phys. Rev. D}
  {\bfseries 90} (2014) 063527},
  [\href{https://arxiv.org/abs/1404.6096}{{\ttfamily 1404.6096}}].

\bibitem{Cognola:2006sp}
G.~Cognola, E.~Elizalde, S.~Nojiri, S.~Odintsov and S.~Zerbini,
  \emph{{String-inspired Gauss-Bonnet gravity reconstructed from the universe
  expansion history and yielding the transition from matter dominance to dark
  energy}}, \href{https://doi.org/10.1103/PhysRevD.75.086002}{\emph{Phys. Rev.
  D} {\bfseries 75} (2007) 086002},
  [\href{https://arxiv.org/abs/hep-th/0611198}{{\ttfamily hep-th/0611198}}].

\bibitem{Ahn:2014fwa}
W.-K. Ahn, B.~Gwak, B.-H. Lee and W.~Lee, \emph{{Instability of Black Holes
  with a Gauss\textendash{}Bonnet Term}},
  \href{https://doi.org/10.1140/epjc/s10052-015-3568-5}{\emph{Eur. Phys. J. C}
  {\bfseries 75} (2015) 372},
  [\href{https://arxiv.org/abs/1412.4189}{{\ttfamily 1412.4189}}].

\bibitem{Khimphun:2016gsn}
S.~Khimphun, B.-H. Lee and W.~Lee, \emph{{Phase transition for black holes in
  dilatonic Einstein-Gauss-Bonnet theory of gravitation}},
  \href{https://doi.org/10.1103/PhysRevD.94.104067}{\emph{Phys. Rev. D}
  {\bfseries 94} (2016) 104067},
  [\href{https://arxiv.org/abs/1605.07377}{{\ttfamily 1605.07377}}].

\bibitem{Lee:2016yaj}
B.-H. Lee, W.~Lee and D.~Ro, \emph{{Fubini instantons in Dilatonic
  Einstein\textendash{}Gauss\textendash{}Bonnet theory of gravitation}},
  \href{https://doi.org/10.1016/j.physletb.2016.09.013}{\emph{Phys. Lett. B}
  {\bfseries 762} (2016) 535--542},
  [\href{https://arxiv.org/abs/1607.01125}{{\ttfamily 1607.01125}}].

\bibitem{Antoniou:2017acq}
G.~Antoniou, A.~Bakopoulos and P.~Kanti, \emph{{Evasion of No-Hair Theorems and
  Novel Black-Hole Solutions in Gauss-Bonnet Theories}},
  \href{https://doi.org/10.1103/PhysRevLett.120.131102}{\emph{Phys. Rev. Lett.}
  {\bfseries 120} (2018) 131102},
  [\href{https://arxiv.org/abs/1711.03390}{{\ttfamily 1711.03390}}].

\bibitem{Doneva:2017bvd}
D.~D. Doneva and S.~S. Yazadjiev, \emph{{New Gauss-Bonnet Black Holes with
  Curvature-Induced Scalarization in Extended Scalar-Tensor Theories}},
  \href{https://doi.org/10.1103/PhysRevLett.120.131103}{\emph{Phys. Rev. Lett.}
  {\bfseries 120} (2018) 131103},
  [\href{https://arxiv.org/abs/1711.01187}{{\ttfamily 1711.01187}}].

\bibitem{Silva:2017uqg}
H.~O. Silva, J.~Sakstein, L.~Gualtieri, T.~P. Sotiriou and E.~Berti,
  \emph{{Spontaneous scalarization of black holes and compact stars from a
  Gauss-Bonnet coupling}},
  \href{https://doi.org/10.1103/PhysRevLett.120.131104}{\emph{Phys. Rev. Lett.}
  {\bfseries 120} (2018) 131104},
  [\href{https://arxiv.org/abs/1711.02080}{{\ttfamily 1711.02080}}].

\bibitem{Myung:2018iyq}
Y.~S. Myung and D.-C. Zou, \emph{{Gregory-Laflamme instability of black hole in
  Einstein-scalar-Gauss-Bonnet theories}},
  \href{https://doi.org/10.1103/PhysRevD.98.024030}{\emph{Phys. Rev. D}
  {\bfseries 98} (2018) 024030},
  [\href{https://arxiv.org/abs/1805.05023}{{\ttfamily 1805.05023}}].

\bibitem{Lee:2018zym}
B.-H. Lee, W.~Lee and D.~Ro, \emph{{Expanded evasion of the black hole no-hair
  theorem in dilatonic Einstein-Gauss-Bonnet theory}},
  \href{https://doi.org/10.1103/PhysRevD.99.024002}{\emph{Phys. Rev. D}
  {\bfseries 99} (2019) 024002},
  [\href{https://arxiv.org/abs/1809.05653}{{\ttfamily 1809.05653}}].

\bibitem{Chew:2020lkj}
X.~Y. Chew, G.~Tumurtushaa and D.-h. Yeom, \emph{{Euclidean wormholes in
  Gauss\textendash{}Bonnet-dilaton gravity}},
  \href{https://doi.org/10.1016/j.dark.2021.100811}{\emph{Phys. Dark Univ.}
  {\bfseries 32} (2021) 100811},
  [\href{https://arxiv.org/abs/2006.04344}{{\ttfamily 2006.04344}}].

\bibitem{Lee:2021uis}
B.-H. Lee, H.~Lee and W.~Lee, \emph{{Hariy black holes in dilatonic
  Einstein-Gauss-Bonnet theory}},  in \emph{{17th Italian-Korean Symposium on
  Relativistic Astrophysics}}, 11, 2021,
  \href{https://arxiv.org/abs/2111.13380}{{\ttfamily 2111.13380}}.

\bibitem{Kawai:2021edk}
S.~Kawai and J.~Kim, \emph{{Primordial black holes from Gauss-Bonnet-corrected
  single field inflation}},
  \href{https://doi.org/10.1103/PhysRevD.104.083545}{\emph{Phys. Rev. D}
  {\bfseries 104} (2021) 083545},
  [\href{https://arxiv.org/abs/2108.01340}{{\ttfamily 2108.01340}}].

\bibitem{Papageorgiou:2022umj}
A.~Papageorgiou, C.~Park and M.~Park, \emph{{Rectifying No-Hair Theorems in
  Gauss-Bonnet theory}},  \href{https://arxiv.org/abs/2205.00907}{{\ttfamily
  2205.00907}}.

\bibitem{Nair:2019iur}
R.~Nair, S.~Perkins, H.~O. Silva and N.~Yunes, \emph{{Fundamental Physics
  Implications for Higher-Curvature Theories from Binary Black Hole Signals in
  the LIGO-Virgo Catalog GWTC-1}},
  \href{https://doi.org/10.1103/PhysRevLett.123.191101}{\emph{Phys. Rev. Lett.}
  {\bfseries 123} (2019) 191101},
  [\href{https://arxiv.org/abs/1905.00870}{{\ttfamily 1905.00870}}].

\bibitem{Okounkova:2020rqw}
M.~Okounkova, \emph{{Numerical relativity simulation of GW150914 in Einstein
  dilaton Gauss-Bonnet gravity}},
  \href{https://doi.org/10.1103/PhysRevD.102.084046}{\emph{Phys. Rev. D}
  {\bfseries 102} (2020) 084046},
  [\href{https://arxiv.org/abs/2001.03571}{{\ttfamily 2001.03571}}].

\bibitem{Wang:2021jfc}
H.-T. Wang, S.-P. Tang, P.-C. Li, M.-Z. Han and Y.-Z. Fan, \emph{{Tight
  constraints on Einstein-dilation-Gauss-Bonnet gravity from GW190412 and
  GW190814}}, \href{https://doi.org/10.1103/PhysRevD.104.024015}{\emph{Phys.
  Rev. D} {\bfseries 104} (2021) 024015}.

\bibitem{Perkins:2021mhb}
S.~E. Perkins, R.~Nair, H.~O. Silva and N.~Yunes, \emph{{Improved
  gravitational-wave constraints on higher-order curvature theories of
  gravity}}, \href{https://doi.org/10.1103/PhysRevD.104.024060}{\emph{Phys.
  Rev. D} {\bfseries 104} (2021) 024060},
  [\href{https://arxiv.org/abs/2104.11189}{{\ttfamily 2104.11189}}].

\bibitem{BH-NS_GB_2022}
Z.~Lyu, N.~Jiang and K.~Yagi, \emph{{Constraints on
  Einstein-dilation-Gauss-Bonnet gravity from black hole-neutron star
  gravitational wave events}},
  \href{https://doi.org/10.1103/PhysRevD.105.064001}{\emph{Phys. Rev. D}
  {\bfseries 105} (2022) 064001},
  [\href{https://arxiv.org/abs/2201.02543}{{\ttfamily 2201.02543}}].

\bibitem{Kusakabe:2015ida}
M.~Kusakabe, S.~Koh, K.~S. Kim and M.-K. Cheoun, \emph{{Constraints on modified
  Gauss-Bonnet gravity during big bang nucleosynthesis}},
  \href{https://doi.org/10.1103/PhysRevD.93.043511}{\emph{Phys. Rev. D}
  {\bfseries 93} (2016) 043511},
  [\href{https://arxiv.org/abs/1507.05565}{{\ttfamily 1507.05565}}].

\bibitem{Asimakis:2021yct}
P.~Asimakis, S.~Basilakos, N.~E. Mavromatos and E.~N. Saridakis, \emph{{Big
  bang nucleosynthesis constraints on higher-order modified gravities}},
  \href{https://doi.org/10.1103/PhysRevD.105.084010}{\emph{Phys. Rev. D}
  {\bfseries 105} (2022) 084010},
  [\href{https://arxiv.org/abs/2112.10863}{{\ttfamily 2112.10863}}].

\bibitem{Amendola:2005cr}
L.~Amendola, C.~Charmousis and S.~C. Davis, \emph{{Constraints on Gauss-Bonnet
  gravity in dark energy cosmologies}},
  \href{https://doi.org/10.1088/1475-7516/2006/12/020}{\emph{JCAP} {\bfseries
  12} (2006) 020}, [\href{https://arxiv.org/abs/hep-th/0506137}{{\ttfamily
  hep-th/0506137}}].

\bibitem{Salati:2002md}
P.~Salati, \emph{{Quintessence and the relic density of neutralinos}},
  \href{https://doi.org/10.1016/j.physletb.2003.07.073}{\emph{Phys. Lett. B}
  {\bfseries 571} (2003) 121--131},
  [\href{https://arxiv.org/abs/astro-ph/0207396}{{\ttfamily
  astro-ph/0207396}}].

\bibitem{Rosati:2003yw}
F.~Rosati, \emph{{Quintessential enhancement of dark matter abundance}},
  \href{https://doi.org/10.1016/j.physletb.2003.07.048}{\emph{Phys. Lett. B}
  {\bfseries 570} (2003) 5--10},
  [\href{https://arxiv.org/abs/hep-ph/0302159}{{\ttfamily hep-ph/0302159}}].

\bibitem{Kang:2008zi}
J.~U. Kang and G.~Panotopoulos, \emph{{Big-Bang Nucleosynthesis and neutralino
  dark matter in modified gravity}},
  \href{https://doi.org/10.1016/j.physletb.2009.05.006}{\emph{Phys. Lett. B}
  {\bfseries 677} (2009) 6--11},
  [\href{https://arxiv.org/abs/0806.1493}{{\ttfamily 0806.1493}}].

\bibitem{Capozziello:2012uv}
S.~Capozziello, M.~De~Laurentis and G.~Lambiase, \emph{{Cosmic relic abundance
  and f(R) gravity}},
  \href{https://doi.org/10.1016/j.physletb.2012.07.007}{\emph{Phys. Lett. B}
  {\bfseries 715} (2012) 1--8},
  [\href{https://arxiv.org/abs/1201.2071}{{\ttfamily 1201.2071}}].

\bibitem{Capozziello:2015ama}
S.~Capozziello, V.~Galluzzi, G.~Lambiase and L.~Pizza, \emph{{Cosmological
  evolution of thermal relic particles in $f(R)$ gravity}},
  \href{https://doi.org/10.1103/PhysRevD.92.084006}{\emph{Phys. Rev. D}
  {\bfseries 92} (2015) 084006},
  [\href{https://arxiv.org/abs/1507.06835}{{\ttfamily 1507.06835}}].

\bibitem{Meehan:2015cna}
M.~T. Meehan and I.~B. Whittingham, \emph{{Dark matter relic density in
  scalar-tensor gravity revisited}},
  \href{https://doi.org/10.1088/1475-7516/2015/12/011}{\emph{JCAP} {\bfseries
  12} (2015) 011}, [\href{https://arxiv.org/abs/1508.05174}{{\ttfamily
  1508.05174}}].

\bibitem{Lambiase:2016log}
G.~Lambiase, \emph{{$f(R)$ cosmology and dark matter}},
  \href{https://doi.org/10.22323/1.268.0012}{\emph{PoS} {\bfseries DSU2015}
  (2016) 012}.

\bibitem{profumo_relentless_2017}
F.~D'Eramo, N.~Fernandez and S.~Profumo, \emph{{When the Universe Expands Too
  Fast: Relentless Dark Matter}},
  \href{https://doi.org/10.1088/1475-7516/2017/05/012}{\emph{JCAP} {\bfseries
  05} (2017) 012}, [\href{https://arxiv.org/abs/1703.04793}{{\ttfamily
  1703.04793}}].

\bibitem{Schelke_2006}
M.~Schelke, R.~Catena, N.~Fornengo, A.~Masiero and M.~Pietroni,
  \emph{{Constraining pre Big-Bang-Nucleosynthesis Expansion using Cosmic
  Antiprotons}}, \href{https://doi.org/10.1103/PhysRevD.74.083505}{\emph{Phys.
  Rev. D} {\bfseries 74} (2006) 083505},
  [\href{https://arxiv.org/abs/hep-ph/0605287}{{\ttfamily hep-ph/0605287}}].

\bibitem{Donato:2006}
F.~Donato, N.~Fornengo and M.~Schelke, \emph{{Additional bounds on the pre
  big-bang-nucleosynthesis expansion by means of gamma-rays from the galactic
  center}}, \href{https://doi.org/10.1088/1475-7516/2007/03/021}{\emph{JCAP}
  {\bfseries 03} (2007) 021},
  [\href{https://arxiv.org/abs/hep-ph/0612374}{{\ttfamily hep-ph/0612374}}].

\bibitem{Buchmuller:2005eh}
W.~Buchmuller, R.~D. Peccei and T.~Yanagida, \emph{{Leptogenesis as the origin
  of matter}},
  \href{https://doi.org/10.1146/annurev.nucl.55.090704.151558}{\emph{Ann. Rev.
  Nucl. Part. Sci.} {\bfseries 55} (2005) 311--355},
  [\href{https://arxiv.org/abs/hep-ph/0502169}{{\ttfamily hep-ph/0502169}}].

\bibitem{Kawai:2017kqt}
S.~Kawai and J.~Kim, \emph{{Gauss\textendash{}Bonnet Chern\textendash{}Simons
  gravitational wave leptogenesis}},
  \href{https://doi.org/10.1016/j.physletb.2018.12.019}{\emph{Phys. Lett. B}
  {\bfseries 789} (2019) 145--149},
  [\href{https://arxiv.org/abs/1702.07689}{{\ttfamily 1702.07689}}].

\bibitem{Boulware:1985wk}
D.~G. Boulware and S.~Deser, \emph{{String Generated Gravity Models}},
  \href{https://doi.org/10.1103/PhysRevLett.55.2656}{\emph{Phys. Rev. Lett.}
  {\bfseries 55} (1985) 2656}.

\bibitem{Choi:2017mkk}
S.-M. Choi, H.~M. Lee and M.-S. Seo, \emph{{Cosmic abundances of SIMP dark
  matter}}, \href{https://doi.org/10.1007/JHEP04(2017)154}{\emph{JHEP}
  {\bfseries 04} (2017) 154},
  [\href{https://arxiv.org/abs/1702.07860}{{\ttfamily 1702.07860}}].

\bibitem{g_T_Steigman2012}
G.~Steigman, B.~Dasgupta and J.~F. Beacom, \emph{{Precise Relic WIMP Abundance
  and its Impact on Searches for Dark Matter Annihilation}},
  \href{https://doi.org/10.1103/PhysRevD.86.023506}{\emph{Phys. Rev. D}
  {\bfseries 86} (2012) 023506},
  [\href{https://arxiv.org/abs/1204.3622}{{\ttfamily 1204.3622}}].

\bibitem{AMS_2014}
{\scshape AMS} collaboration, M.~Aguilar et~al., \emph{{Electron and Positron
  Fluxes in Primary Cosmic Rays Measured with the Alpha Magnetic Spectrometer
  on the International Space Station}},
  \href{https://doi.org/10.1103/PhysRevLett.113.121102}{\emph{Phys. Rev. Lett.}
  {\bfseries 113} (2014) 121102}.

\bibitem{AMS_2019}
{\scshape AMS} collaboration, M.~Aguilar et~al., \emph{{Towards Understanding
  the Origin of Cosmic-Ray Positrons}},
  \href{https://doi.org/10.1103/PhysRevLett.122.041102}{\emph{Phys. Rev. Lett.}
  {\bfseries 122} (2019) 041102}.

\bibitem{AMS_DM_2017}
L.~A. Cavasonza, H.~Gast, M.~Krämer, M.~Pellen and S.~Schael,
  \emph{Constraints on leptophilic dark matter from the ams-02 experiment},
  \href{https://doi.org/10.3847/1538-4357/aa624d}{\emph{The Astrophysical
  Journal} {\bfseries 839} (apr, 2017) 36}.

\bibitem{Fermi-LAT_2017GC}
{\scshape Fermi-LAT} collaboration, M.~Ackermann et~al., \emph{{The Fermi
  Galactic Center GeV Excess and Implications for Dark Matter}},
  \href{https://doi.org/10.3847/1538-4357/aa6cab}{\emph{Astrophys. J.}
  {\bfseries 840} (2017) 43},
  [\href{https://arxiv.org/abs/1704.03910}{{\ttfamily 1704.03910}}].

\bibitem{Fermi-LAT_2017}
{\scshape Fermi-LAT, DES} collaboration, A.~Albert et~al., \emph{{Searching for
  Dark Matter Annihilation in Recently Discovered Milky Way Satellites with
  Fermi-LAT}},
  \href{https://doi.org/10.3847/1538-4357/834/2/110}{\emph{Astrophys. J.}
  {\bfseries 834} (2017) 110},
  [\href{https://arxiv.org/abs/1611.03184}{{\ttfamily 1611.03184}}].

\bibitem{CMB_2015}
T.~R. Slatyer, \emph{{Indirect dark matter signatures in the cosmic dark ages.
  I. Generalizing the bound on s-wave dark matter annihilation from Planck
  results}}, \href{https://doi.org/10.1103/PhysRevD.93.023527}{\emph{Phys. Rev.
  D} {\bfseries 93} (2016) 023527},
  [\href{https://arxiv.org/abs/1506.03811}{{\ttfamily 1506.03811}}].

\bibitem{pppc_2010}
M.~Cirelli, G.~Corcella, A.~Hektor, G.~Hutsi, M.~Kadastik, P.~Panci et~al.,
  \emph{{PPPC 4 DM ID: A Poor Particle Physicist Cookbook for Dark Matter
  Indirect Detection}},
  \href{https://doi.org/10.1088/1475-7516/2011/03/051}{\emph{JCAP} {\bfseries
  03} (2011) 051}, [\href{https://arxiv.org/abs/1012.4515}{{\ttfamily
  1012.4515}}].

\bibitem{indirect_search_Beacom_2018}
R.~K. Leane, T.~R. Slatyer, J.~F. Beacom and K.~C.~Y. Ng, \emph{Gev-scale
  thermal wimps: Not even slightly ruled out},
  \href{https://doi.org/10.1103/PhysRevD.98.023016}{\emph{Phys. Rev. D}
  {\bfseries 98} (Jul, 2018) 023016}.

\bibitem{planck_2018}
{\scshape Planck} collaboration, N.~Aghanim et~al., \emph{{Planck 2018 results.
  VI. Cosmological parameters}},
  \href{https://doi.org/10.1051/0004-6361/201833910}{\emph{Astron. Astrophys.}
  {\bfseries 641} (2020) A6},
  [\href{https://arxiv.org/abs/1807.06209}{{\ttfamily 1807.06209}}].

\bibitem{Sotiriou:2006pq}
T.~P. Sotiriou and E.~Barausse, \emph{{Post-Newtonian expansion for
  Gauss-Bonnet gravity}},
  \href{https://doi.org/10.1103/PhysRevD.75.084007}{\emph{Phys. Rev. D}
  {\bfseries 75} (2007) 084007},
  [\href{https://arxiv.org/abs/gr-qc/0612065}{{\ttfamily gr-qc/0612065}}].

\bibitem{Yagi:2015oca}
K.~Yagi, L.~C. Stein and N.~Yunes, \emph{{Challenging the Presence of Scalar
  Charge and Dipolar Radiation in Binary Pulsars}},
  \href{https://doi.org/10.1103/PhysRevD.93.024010}{\emph{Phys. Rev. D}
  {\bfseries 93} (2016) 024010},
  [\href{https://arxiv.org/abs/1510.02152}{{\ttfamily 1510.02152}}].

\bibitem{Yagi:2011xp}
K.~Yagi, L.~C. Stein, N.~Yunes and T.~Tanaka, \emph{{Post-Newtonian,
  Quasi-Circular Binary Inspirals in Quadratic Modified Gravity}},
  \href{https://doi.org/10.1103/PhysRevD.85.064022}{\emph{Phys. Rev. D}
  {\bfseries 85} (2012) 064022},
  [\href{https://arxiv.org/abs/1110.5950}{{\ttfamily 1110.5950}}].

\bibitem{Witek:2018dmd}
H.~Witek, L.~Gualtieri, P.~Pani and T.~P. Sotiriou, \emph{{Black holes and
  binary mergers in scalar Gauss-Bonnet gravity: scalar field dynamics}},
  \href{https://doi.org/10.1103/PhysRevD.99.064035}{\emph{Phys. Rev. D}
  {\bfseries 99} (2019) 064035},
  [\href{https://arxiv.org/abs/1810.05177}{{\ttfamily 1810.05177}}].

\bibitem{Blazquez-Salcedo:2016enn}
J.~L. Bl\'azquez-Salcedo, C.~F.~B. Macedo, V.~Cardoso, V.~Ferrari,
  L.~Gualtieri, F.~S. Khoo et~al., \emph{{Perturbed black holes in
  Einstein-dilaton-Gauss-Bonnet gravity: Stability, ringdown, and
  gravitational-wave emission}},
  \href{https://doi.org/10.1103/PhysRevD.94.104024}{\emph{Phys. Rev. D}
  {\bfseries 94} (2016) 104024},
  [\href{https://arxiv.org/abs/1609.01286}{{\ttfamily 1609.01286}}].

\bibitem{LIGOScientific:2018mvr}
{\scshape LIGO Scientific, Virgo} collaboration, B.~P. Abbott et~al.,
  \emph{{GWTC-1: A Gravitational-Wave Transient Catalog of Compact Binary
  Mergers Observed by LIGO and Virgo during the First and Second Observing
  Runs}}, \href{https://doi.org/10.1103/PhysRevX.9.031040}{\emph{Phys. Rev. X}
  {\bfseries 9} (2019) 031040},
  [\href{https://arxiv.org/abs/1811.12907}{{\ttfamily 1811.12907}}].

\end{thebibliography}

\end{document}